\input harvmac



\def\unlockat{\catcode`\@=11}
\def\lockat{\catcode`\@=12}

\unlockat

\def\newsec#1{\global\advance\secno by1\message{(\the\secno. #1)}
\global\subsecno=0\global\subsubsecno=0\eqnres@t\noindent
{\bf\the\secno. #1}
\writetoca{{\secsym} {#1}}\par\nobreak\medskip\nobreak}
\global\newcount\subsecno \global\subsecno=0
\def\subsec#1{\global\advance\subsecno
by1\message{(\secsym\the\subsecno. #1)}
\ifnum\lastpenalty>9000\else\bigbreak\fi\global\subsubsecno=0
\noindent{\it\secsym\the\subsecno. #1}
\writetoca{\string\quad {\secsym\the\subsecno.} {#1}}
\par\nobreak\medskip\nobreak}
\global\newcount\subsubsecno \global\subsubsecno=0
\def\subsubsec#1{\global\advance\subsubsecno by1
\message{(\secsym\the\subsecno.\the\subsubsecno. #1)}
\ifnum\lastpenalty>9000\else\bigbreak\fi
\noindent\quad{\secsym\the\subsecno.\the\subsubsecno.}{#1}
\writetoca{\string\qquad{\secsym\the\subsecno.\the\subsubsecno.}{#1}}
\par\nobreak\medskip\nobreak}

\def\subsubseclab#1{\DefWarn#1\xdef
#1{\noexpand\hyperref{}{subsubsection}%
{\secsym\the\subsecno.\the\subsubsecno}%
{\secsym\the\subsecno.\the\subsubsecno}}%
\writedef{#1\leftbracket#1}\wrlabeL{#1=#1}}
\lockat

\def\IL{\relax{\rm I\kern-.18em L}}
\def\IH{\relax{\rm I\kern-.18em H}}
\def\IR{\relax{\rm I\kern-.18em R}}
\def\IC{\relax\hbox{$\inbar\kern-.3em{\rm C}$}}
\def\IT{\relax\hbox{$\inbar\kern-.3em{\rm T}$}}
\def\IZ{\relax\ifmmode\mathchoice
{\hbox{\cmss Z\kern-.4em Z}}{\hbox{\cmss Z\kern-.4em Z}}
{\lower.9pt\hbox{\cmsss Z\kern-.4em Z}}
{\lower1.2pt\hbox{\cmsss Z\kern-.4em Z}}\else{\cmss Z\kern-.4em
Z}\fi}
\def\CM {{\cal M}}

\def\CF {{\cal F}}

\def\CH {{\cal H}}

\def\bs2{{\bf S}^2}

\def\CM {{\cal M}}


\def\half {{1\over 2}}

\def\CM {{\cal M}}


\def\IB{\relax{\rm I\kern-.18em B}}
\def\IC{{\relax\hbox{$\inbar\kern-.3em{\rm C}$}}}
\def\ID{\relax{\rm I\kern-.18em D}}
\def\IE{\relax{\rm I\kern-.18em E}}
\def\IF{\relax{\rm I\kern-.18em F}}
\def\IG{\relax\hbox{$\inbar\kern-.3em{\rm G}$}}
\def\IGa{\relax\hbox{${\rm I}\kern-.18em\Gamma$}}
\def\IH{\relax{\rm I\kern-.18em H}}
\def\II{\relax{\rm I\kern-.18em I}}
\def\IK{\relax{\rm I\kern-.18em K}}
\def\IP{\relax{\rm I\kern-.18em P}}
\def\IQ{\relax\hbox{$\inbar\kern-.3em{\rm Q}$}}

\def\inbar{\,\vrule height1.5ex width.4pt depth0pt}

\def\mod{{\rm mod}}

\font\cmss=cmss10 \font\cmsss=cmss10 at 7pt
\def\IR{\relax{\rm I\kern-.18em R}}

\def\vol{{\rm vol}}


\def\boxit#1{\vbox{\hrule\hbox{\vrule\kern8pt
\vbox{\hbox{\kern8pt}\hbox{\vbox{#1}}\hbox{\kern8pt}}
\kern8pt\vrule}\hrule}}
\def\mathboxit#1{\vbox{\hrule\hbox{\vrule\kern8pt\vbox{\kern8pt
\hbox{$\displaystyle #1$}\kern8pt}\kern8pt\vrule}\hrule}}


\def\inbar{\,\vrule height1.5ex width.4pt depth0pt}

\font\cmss=cmss10 \font\cmsss=cmss10 at 7pt
\def\IR{\relax{\rm I\kern-.18em R}}

\def\vol{{\rm vol}}

\def\frac#1#2{{{#1}\over{#2}}}
\def\half{{1\over2}}
\def\raiz{\sqrt{2}}
\def\ex{{\hbox{\rm e}}}

\def\gof{\Gamma^0(4)}



\lref\DoKro{S.K. Donaldson and P.B. Kronheimer, {\it
The geometry of four-manifolds}, Oxford, 1990.}

\lref\borchaut{R.E. Borcherds,
``Automorphic forms with singularities on
Grassmannians,'' Invent. Math. {\bf 132} (1998) 491.}

\lref\gottsche{L. G\"ottsche, ``Modular forms and Donaldson
invariants for 4-manifolds with $b_+=1$,'' alg-geom/9506018; J. Am. Math. Soc.
{\bf 9}
(1996) 827.}

\lref\gottzag{L. G\"ottsche and D. Zagier,
``Jacobi forms and the structure of Donaldson
invariants for 4-manifolds with $b_+=1$,''
alg-geom/9612020, Sel. math., New ser. {\bf 4} (1998) 69.}

\lref\mw{G. Moore and E. Witten, ``Integration over
the $u$-plane in Donaldson theory," hep-th/9709193, Adv. Theor. Math. Phys.
{\bf 1} (1997) 298.}

\lref\swi{N. Seiberg and E. Witten,
``Electric-magnetic duality, monopole condensation, and confinement in
${\cal N}=2$ supersymmetric Yang-Mills
theory,''
hep-th/9407087, Nucl. Phys. {\bf B 426} (1994) 19.}

\lref\swii{N. Seiberg and E. Witten,
``Monopoles, duality and chiral symmetry breaking in ${\cal N}=2$
supersymmetric
QCD,''
hep-th/9408099, Nucl. Phys. {\bf B 431} (1994) 484.}

\lref\mfm{E. Witten, ``Monopoles and
four-manifolds,''  hep-th/9411102; Math. Res. Letters {\bf 1} (1994) 769.}

%
%

\lref\wittk{E. Witten, ``Supersymmetric Yang-Mills theory
on a four-manifold,''  hep-th/9403193;
J. Math. Phys. {\bf 35} (1994) 5101.}

\lref\zag{D. Zagier, ``On the cohomology of moduli spaces of
rank two vector bundles over curves," in {\it The moduli space of curves},
R. Dijkgraaf {\it et al.} eds., Birkh\"auser.}

\lref\news{P.E. Newstead, ``Characteristic classes of stable bundles over an
algebraic curve," Trans. Amer. Math. Soc. {\bf 169} (1972) 337.}

\lref\lns{A. Losev, N. Nekrasov, and S. Shatashvili, ``Issues in
topological gauge theory," hep-th/9711108,
Nucl. Phys. {\bf B 549} (1998); ``Testing Seiberg-Witten solution,"
hep-th/9801061.}

\lref\mmone{M. Mari\~no and G. Moore, ``Integrating over the Coulomb branch in
${\cal N}=2$ gauge theory," hep-th/9712062, Nucl. Phys. {\bf B} (Proc.
Suppl) {\bf 68}
(1998) 336; ``The Donaldson-Witten function for gauge
groups
of rank larger than one," hep-th/9802185, Commun. Math. Phys. {\bf 199}
(1998) 25.}

\lref\mmtwo{M. Mari\~no and G. Moore, ``Donaldson invariants for non-simply
connected
manifolds," hep-th/9804114, Commun. Math. Phys. {\bf 203} (1999) 249.}

\lref\lalo{J.M.F. Labastida and C. Lozano, ``Duality in twisted
${\cal N}=4$
supersymmetric gauge theories in four dimensions," hep-th/9806032, Nucl.
Phys. {\bf
B 537} (1999) 203; H. Kanno and S.-K. Yang,
``Donaldson-Witten functions of massless ${\cal N}=2$ supersymmetric QCD,"
hep-th/9806015,
Nucl. Phys. {\bf B 535} (1998) 512.}

\lref\mmthree{M. Mari\~no and G. Moore, ``Three-manifold topology and the
Donaldson-Witten partition function," hep-th/9811214, Nucl. Phys. {\bf B
547} (1999) 569.}

\lref\mmp{M. Mari\~no, G. Moore and G. Peradze, ``Superconformal invariance
and the
geography of four-manifolds," hep-th/9812055; ``Four-manifold geography and
superconformal
symmetry," math.DG/9812042.}

\lref\whith{A. Gorsky, A. Marshakov, A. Mironov and A.Morozov, ``RG equations
from
Whitham hierarchy," hep-th/9802007, Nucl. Phys. {\bf B 527} (1998) 690;
K. Takasaki, ``Integrable hierarchies and contact terms in $u$-plane
integrals of
topologically twisted supersymmetric gauge theories," hep-th/9803217, Int.
J. Mod. Phys. {\bf A 14}
(1999) 1001; M. Mari\~no, ``The uses of Whitham hierarchies," hep-th/9905053.}

\lref\tqft{E. Witten,
``Topological Quantum Field Theory,''
Commun. Math. Phys. {\bf 117} (1988)
353.}

\lref\ab{M.F. Atiyah and R. Bott, ``The Yang-Mills equations
over Riemann surfaces," Philos. Trans. Roy. Soc. London {\bf 308} (1982)
523.}

\lref\th{M. Thaddeus, ``Conformal field theory and the cohomology of the
moduli space
of stable bundles," J. Diff. Geom. {\bf 35} (1992) 131.}

\lref\bt{M. Blau and G. Thompson, ``Lectures on 2d gauge theories,"
hep-th/9310144, in {\it 1993 Trieste Summer School in High Energy
Physics and Cosmology},
E. Gava et al., eds., World Scientific, 1994.}

\lref\bjsv{ M. Bershadsky, A. Johansen, V. Sadov, and
C. Vafa, ``Topological Reduction of 4D SYM to 2D $\sigma$--Models,''
hep-th/9501096; Nucl. Phys. {\bf B448} (1995) 166.}

\lref\donaldsonii{S.K. Donaldson, ``Floer homology and
algebraic geometry,'' in {\it Vector bundles in
algebraic geometry}, N.J. Hitchin et. al. eds.,
Cambridge University Press, 1995.}

\lref\floer{{\it The Floer Memorial Volume}, H. Hofer et. al.
eds., Birkh\"auser 1995.}

\lref\munoz{V. Mu\~noz, ``Wall-crossing formulae for algebraic surfaces with
positive irregularity," alg-geom/9709002.}

\lref\munozfloer{V. Mu\~noz, ``Ring structure of the Floer cohomology of
$\Sigma \times {\bf S}^1$," dg-ga/9710029, Topology {\bf 38} (1999) 517.}

\lref\munozff{V. Mu\~noz, ``Fukaya-Floer homology of $\Sigma \times {\bf
S}^1$ and
applications," math.DG/9804081}

\lref\munozapp{V. Mu\~noz, ``Basic classes for four-manifolds not of simple
type,"
math.DG/9811089; ``Higher type adjunction inequalities for Donaldson
invariants,"
math.DG/9901046.}

\lref\munozqu{V. Mu\~noz, ``Quantum cohomology of the moduli space of stable
bundles over a Riemann surface," alg-geom/9711030, Duke Math. J. {\bf 98}
(1999) 525.}

\lref\okonek{T.J. Li and A. Liu, ``General wall-crossing formula,'' Math. Res.
Lett. {\bf 2} (1995) 797; C. Okonek and A. Teleman, ``Seiberg-Witten
invariants for
manifolds with $b_2^+=1$, and the universal wall-crossing formula,"
alg-geom/9603003, Int. J. Math. {\bf 7} (1996) 811.}

\lref\froy{K.A. Froyshov, ``Equivariant aspects of Yang-Mills-Floer theory,"
 math.DG/9903083.}

\lref\harvmoore{L. Dixon, V. Kaplunovsky and J. Louis, ``Moduli dependence
of string loop corrections
to gauge coupling constants," Nucl. Phys. {\bf B 355} (1991) 649. J.A.
Harvey and G. Moore,
``Algebras, BPS states, and strings," hep-th/9510182, Nucl. Phys. {\bf B
463} (1996) 315.}

\lref\verlinde{E. Verlinde, ``Fusion rules and modular transformations in 2d
conformal field theory," Nucl. Phys. {\bf B 300} (1988) 360. }

\lref\msz{J. Morgan and Z. Szab\'o, ``Embedded tori in four-manifolds,"
Topology {\bf 38} (1999) 479.}

\lref\donglu{S.K. Donaldson, ``Gluing techniques in the cohomology of
moduli spaces," in {\it
Topological methods in modern mathematics}, L.R. Goldberg and A.V. Phillips
eds., Publish or Perish,
1993.}

\lref\qtwod{E. Witten, ``On quantum gauge theories in two dimensions,"
Commun. Math. Phys. {\bf 141}
(1991) 153.}

\lref\revis{E. Witten, ``Two-dimensional gauge theories  revisited,"
hep-th/9204083,
J. Geom. Phys. {\bf 9} (1992) 303.}

\lref\at{M.F. Atiyah, {\it The geometry and physics of knots}, Cambridge
University Press, 1990; R. Dijkgraaf, ``Fields, strings and
duality," hep-th/9703136, in {\it Symm\'etries quantiques} (Les
Houches, 1995), North-Holland, Amsterdam, 1998.}

\lref\ds{S. Dostoglou and D. Salamon, ``Self-dual instantons and
holomorphic curves,"
Annals of Mathematics {\bf 139} (1994) 581.}

\lref\st{B. Siebert and G. Tian, ``Recursive relations for the cohomology ring
of moduli spaces of stable bundles," alg-geom/9410019,
Turkish J. Math. {\bf 19} (1995) 131;
A.D. King and P.E. Newstead, ``On the cohomology ring of the
moduli space
of stable bundles on a curve," Topology {\bf 37} (1998) 407.}

\lref\sieb{B. Siebert, ``An update on (small) quantum cohomology," in {\it
Mirror Symmetry III}, AMS,
1999. }

\lref\kiem{Y.-H. Kiem, ``Equivariant and intersection cohomology of moduli
spaces of vector bundles"; ``Intersection cohomology of quotients of
non-singular
varieties," Yale preprints.}

\lref\hatter{L. Hatter, ``Cohomology of compactifications of moduli spaces
of stable
bundles over a Riemann surface," Ph.D. Thesis, Oxford University, 1997
(unpublished).}

\lref\szenes{A. Szenes, ``The combinatorics of the Verlinde formulas,"
alg-geom/9402003,
in {\it Vector bundles in algebraic geometry}, N.J. Hitchin et. al. eds.,
Cambridge University Press, 1995.}

\lref\mohri{K. Mohri, ``Residues and topological Yang-Mills theory in two
dimensions,"
hep-th/9604022, Rev. Math. Phys. {\bf 9} (1997) 59.}


\Title{\vbox{\baselineskip12pt
\hbox{US-FT/16-99}
\hbox{YCTP-P18-99 }
\hbox{hep-th/9907165}
}}
{\vbox{\centerline{Donaldson invariants of product ruled surfaces}
\centerline{ }
\centerline{and two-dimensional gauge theories}}
}
\centerline{Carlos Lozano$^{a}$ and Marcos Mari\~no$^{b}$}

\bigskip
\medskip
{\vbox{\centerline{$^{a}$ \sl Departamento de F\'\i sica de Part\'\i culas}
\centerline{\sl Universidade de Santiago de Compostela}
\vskip2pt
\centerline{\sl E-15706 Santiago de Compostela, Spain}}
\centerline{ \it lozano@fpaxp1.usc.es}

\bigskip
\medskip
{\vbox{\centerline{$^{b}$ \sl Department of Physics, Yale University}
\vskip2pt
\centerline{\sl New Haven, CT 06520, USA }}
\centerline{ \it marcos.marino@yale.edu }

\bigskip
\bigskip
\noindent
Using the $u$-plane integral of Moore and Witten, we derive a simple
expression for the Donaldson invariants of product ruled surfaces
$\Sigma_g \times {\bf S}^2$, where $\Sigma_g$ is a Riemann surface
of genus $g$. This expression generalizes a theorem of Morgan and
Szab\'o for $g=1$ to any genus $g$. We give two applications of our
results: (1) We derive Thaddeus' formulae for the intersection
pairings on the moduli space of rank two stable bundles over a
Riemann surface. (2) We derive the eigenvalue spectrum of the
Fukaya-Floer cohomology of $\Sigma_g \times {\bf S}^1$.

\Date{June, 1999}

\newsec{Introduction}

The Donaldson invariants of smooth four-manifolds have played a
very important role in physics and mathematics in the last
years. Since the reformulation of Donaldson theory by Witten in terms
of twisted ${\cal N}=2$ Yang-Mills theory \tqft,
the physical approach to Donaldson theory has opened unsuspected
perspectives. The main breakthrough, in this respect, was the
introduction of Seiberg-Witten invariants in \mfm\ and Witten's
``magic" formula relating the Donaldson and Seiberg-Witten invariants
of simply-connected four-manifolds with $b_2^+>1$ and of simple type.
This relation was fully explained in the fundamental paper of Moore
and Witten \mw, which also analyzed Donaldson-Witten theory for
manifolds of $b_2^+=1$ using the formalism of the $u$-plane integral.

The $u$-plane integral of Moore and Witten has been studied
during the last two years from many different points of view
\lns\mmone\mmtwo\lalo\mmthree\mmp\whith. One of the most interesting outcomes
of this approach has been a complete understanding of Donaldson
invariants of non-simply connected manifolds. The study of this
problem from the point of view of the $u$-plane integral was
initiated in \mw\lns\ and completed in \mmtwo, where (among other
things) a general wall-crossing formula for non-simply connected
manifolds was derived, generalizing the results obtained in \munoz\
using algebro-geometric methods.

\subsec{Product ruled surfaces}

Among non-simply connected manifolds of $b_2^+=1$, product ruled
surfaces play an important role in Donaldson theory. These are
four-manifolds of the form $\Sigma_g \times {\bf S}^2$, where
$\Sigma_g$ is a Riemann surface of genus $g$. In \mmtwo, a direct
application of the lattice reduction technique of \mw\ led to
explicit expressions for the Donaldson invariants of these
surfaces, in the chamber where the volume of ${\bf S}^2$ is small.
Using these expressions and summing up the infinite number of
wall-crossing terms, one can derive in principle the
Donaldson-Witten generating function of product ruled surfaces in
the chamber where the volume of $\Sigma_g$ is small. This was the
approach followed in \mmtwo, where explicit and compact formulae
were written down using Kronecker's double identity to sum up the
wall-crossing terms.

However, formulae for the Donaldson invariants based on wall-crossing
tend to be ineffective when the instanton number is large. Based on physical
intuition, we would expect that some of the properties of the Donaldson-Witten
series will not be apparent in these kinds of expressions, which are based
on calculations made in the ``electric" frame. The
approach based on wall-crossing
formulae makes it difficult to write generating formulae,
even for lower genus,
and in fact one of the motivations of this paper was to reproduce the
simple result of Morgan and Szab\'o for $g=1$ \msz\ using the $u$-plane
integral.

The approach that we follow in this paper is to perform a direct calculation of
the $u$-plane integral in the chamber where $\Sigma_g$ is very small. This
requires
a slight generalization of the computation in \mw\ to allow for a non-zero
Stiefel-Whitney class. In this chamber, there is a very important
contribution coming from  the Seiberg-Witten invariants. In this case, this
contribution can be also
computed from the $u$-plane integral via
wall-crossing, with the important difference that
the number of walls is always finite. This is the main reason for the relative
simplicity of our final expression, which is expressed in terms of
``magnetic" variables.
Therefore, together with the results given in \mmtwo, one finds two different
expressions for the Donaldson invariants of product ruled surfaces. This is
somewhat
similar to the case of ${\IC P}^2$, whose invariants were computed in
\gottsche\ by summing
up an infinite number of wall-crossings, and in \mw\ by direct evaluation
of the $u$-plane
integral. There is, however, one important difference: in the case of
${\IC P}^2$ both expressions are expressed in terms of ``electric"
variables, while in
this case one of them is written in ``electric" variables, and the other in
``magnetic" variables. Depending on the problem we are interested on,
we will find useful one expression
or the other.

\subsec{Relation to the moduli space of stable bundles on a curve}

Apart from its intrinsic interest, the importance of having
explicit expressions for the Donaldson invariants of product ruled
surfaces comes from their relation to other interesting moduli
problems. First of all, for zero instanton number, the moduli space
of instantons on $\Sigma_g\times {\bf S}^2$ is nothing but the
moduli space of flat connections on the Riemann surface $\Sigma_g$.
This space has been extensively studied by mathematicians, and the
structure of its cohomology ring has been explored using
gauge-theoretic techniques, starting with the seminal paper of
Atiyah and Bott \ab. Using the connection to Verlinde formula
\verlinde, Thaddeus \th\ was able to compute the intersection
pairings for the generators of the cohomology ring. The moduli
space of flat connections can be also described by a
two-dimensional version of Donaldson-Witten theory. In \revis,
Witten gave a physical derivation of these intersection pairings by
exploiting the relation of this two-dimensional topological field
theory to physical 2d Yang-Mills theory \qtwod. He found in fact explicit
formulae
for higher
rank gauge groups. In this paper, we will give another derivation
of Thaddeus' formulae using the Donaldson invariants of product
ruled surfaces. In a sense, our derivation can be regarded as the
dimensional reduction of Donaldson-Witten theory down to two
dimensions. In this case, as we are considering zero instanton
number, it is preferable to use the ``electric" expressions given in
\mmtwo.

\subsec{Relation to Fukaya-Floer cohomology}

Another important application of the Donaldson invariants of product ruled
surfaces
is to the (Fukaya)-Floer cohomology of $\Sigma_g \times {\bf S}^1$. The
Fukaya-Floer cohomology
of $\Sigma_g \times {\bf S}^1$ gives the gluing theory for Donaldson
invariants along
this three-manifold, and the gluing theory can be used in turn to derive
important
properties of Donaldson invariants of general four-manifolds
\munozff\munozapp\froy. The ring
structure of the Floer cohomology of $\Sigma_g \times {\bf S}^1$ has been
studied from many different
points of view. In \bjsv, an explicit presentation was obtained under the
assumption that
the eigenvalues can be obtained from the Donaldson invariants of
$\Sigma_g\times \Sigma_1$.
This presentation was finally derived by V. Mu\~noz in \munozfloer, and a
partial determination
of the ring structure of the Fukaya-Floer cohomology was obtained in \munozff.
In a sense, the information contained in the Fukaya-Floer cohomology of
$\Sigma_g \times {\bf S}^1$
is equivalent to the information contained in the Donaldson invariants of
product ruled surfaces.
Using our ``magnetic" expression for the Donaldson invariants, it is
straightforward to find the
eigenvalue spectrum of the Fukaya-Floer cohomology. It is interesting to
notice that this spectrum
is by no means obvious from the ``electric" expressions,
in other words, it can not be seen in a semiclassical
instanton expansion.

\subsec{Relation to the quantum cohomology of the moduli space of stable
bundles over curves}

We have seen that the Donaldson invariants of product ruled
surfaces that correspond to zero Pontriagin number give the
intersection pairings on the moduli space of stable bundles over a
Riemann surface. It has been argued that the Donaldson invariants
with non-zero Pontriagin numbers, when computed in the chamber where
$\Sigma_g$ is small and with Stiefel-Whitney classes satisfying
$(w_2(E), [\Sigma_g]) \not=0$, give essentially the Gromov-Witten
invariants of this moduli space. This was shown in \bjsv\ using the
dimensional reduction of topological Yang-Mills theory on $\Sigma_g
\times {\bf S}^2$ to a type A topological sigma model whose target space
was the moduli space of flat connections on the Riemann surface
\foot{In \lns\ it was argued,
by performing the dimensional reduction in the low-energy action, that
the effective two-dimensional theory can be formulated
in terms of a topological Landau-Ginzburg model, which would then give an
equivalent description of the quantum cohomology in a way reminiscent of
mirror symmetry. It would be interesting
to check this in some detail.}.
The relation between the Gromov-Witten invariants and the
Donaldson invariants is equivalent to the Atiyah-Floer conjecture,
which says that the Floer cohomology of $\Sigma_g \times {\bf S}^1$
is isomorphic as a ring to the quantum cohomology of the moduli space
of stable bundles over $\Sigma_g$. The isomorphism of vector spaces
was proved in \ds, and the ring isomorphism was proved in \munozqu.
Using this isomorphism, one can interpret our formula (5.15) for the
generating function of
the Donaldson invariants as a generating function for the Gromov-Witten
invariants.

\subsec{Organization of the paper}

The organization of this paper is as follows: in section 2, we give a brief
summary of the
results of \mmtwo\ for the Donaldson invariants of non-simply connected
manifolds. In section
3, we compute the Donaldson invariants of product ruled surfaces in the
chambers of small
volume for $\Sigma_g$ and for ${\bf S}^2$ by direct evaluation. In section
4, we derive
Thaddeus' formulae
for the intersection pairing and Verlinde's formulae. In section 5, we
explain the connection
to Fukaya-Floer cohomology and derive the eigenvalue spectrum.

\newsec{Donaldson-Witten theory on non-simply connected manifolds}

The Donaldson invariants of smooth, compact, oriented four-manifolds
$X$ \DoKro\ are defined by using
intersection theory on the moduli space of anti-self-dual connections. The
cohomology classes
on this space are associated to homology classes of $X$ through the slant
product \DoKro\ or, in the
context of topological field theory, by using
the descent procedure \tqft. In this paper, we will restrict ourselves to
the Donaldson invariants
associated to zero, one and two-homology classes. The inclusion of
three-classes has been
considered in \mmtwo. Define
\eqn\ax{
{\bf A}(X)={\rm Sym}(H_0(X) \oplus H_2(X))\otimes \wedge ^* H_1(X).}
Then, the Donaldson invariants can be regarded as functionals
\eqn\poly{
{\cal D}^{w_2(E)}_X: {\bf A}(X) \rightarrow \IQ, }
where $w_2(E) \in H^2(X, {\bf Z})$ is the second Stiefel-Whitney class
of the gauge bundle. It is convenient to organize these invariants as follows.
Let $\{\delta_i\}_{i=1,\ldots,b_1}$ be a basis of one-cycles,
$\{\beta_i\}_{i=1,\ldots,b_1}$ the corresponding dual basis of
harmonic one-forms, and
$\{S_i\}_{i=1,\ldots, b_2}$ a basis of two-cycles. We introduce
the formal sums
\eqn\cycles{
\delta= \sum_{i=1}^{b_1} \zeta_i \,  \delta_i , \qquad\qquad S=
\sum_{j=1}^{b_2}
v_i \,S_i,}
where  $v_i$ are complex
numbers, and $\zeta_i$ are Grassmann variables. The generator of the
$0$-class will
be denoted by $x \in H_0(X, \IZ)$. We then define the Donaldson-Witten
generating function:
\eqn\donwi{
Z_{DW}(p, \zeta_i,v_i)={\cal D}_X^{w_2(E)}({\rm e}^{p x + \delta + S})
}
so that the Donaldson invariants can be read off from the expansion of the
left-hand side in powers of $p$, $\zeta_i$ and $v_i$. The main result in \tqft\
is that $Z_{DW}$ can be understood as the generating
functional  of a twisted version of
the ${\cal N}=2$ supersymmetric gauge theory -- with gauge group $SU(2)$ -- in
four dimensions -- see \tqft\mw\wittk\ for details.
In the twisted theory one can define observables ${\cal O}(x)$, $
I_1(\delta)=\int_{\delta} {\cal O}_{1}$,
$I_2(S)=\int_{S}{\cal O}_{2}$ (where
${\cal O}_i$ are functionals  of the fields of the theory) in one to
one correspondence with the homology classes of $X$, and in such a
way that the generating functional
$$
\langle \ex^{p{\cal O}+ I_1(\delta)+ I_2(S)}\rangle,
$$
is precisely $Z_{DW}(p,\zeta_i,v_i)$.

Based on the low-energy effective
descriptions of ${\cal N}=2$ gauge theories obtained in \swi\swii, Witten
obtained
a explicit formula for \donwi\ in terms of Seiberg-Witten invariants for
manifolds
of $b_2^+>1$ and simple type \mfm. The general framework to give a complete
evaluation
of \donwi\ was established in \mw. The main result of Moore and Witten is
an explicit
expression for the generating function
$Z_{DW}$:
\eqn\donwii{
Z_{DW}=Z_u+Z_{SW}
}
which consists of two pieces. $Z_{SW}$ is the contribution from the moduli
space ${\cal M}_{SW}$ of solutions of the Seiberg-Witten monopole equations.
$Z_u$ (the $u$-plane integral henceforth) is the integral of a
certain modular form over the fundamental domain of the group $\Gamma^{0}(4)$,
that is, over the quotient $\Gamma^{0}(4)\setminus{\cal H}$,
where ${\cal H}$ is the upper half-plane. The explicit form of $Z_u$ was
derived
in \mw\ for simply connected four-manifolds, and extended to the
non-simply connected case in \mmtwo. $Z_u$ is non-vanishing only for
manifolds with $b_2^{+}=1$, and provides a simple physical explanation of the
failure of topological invariance of the Donaldson invariants on those
manifolds \mw.

\subsec{The $u$-plane integral}

We will start by considering the $u$-plane piece. We will assume for simplicity
that $b_1$ is even, although the general story is very similar.
We can assume that $X$ has $b_2^+=1$ (otherwise the $u$-plane integral is
zero). In this case, there
is a normalized self-dual two form or {\sl period point} $\omega$, with
$\omega^2=1$, which
generates $H^{2,+}(X,\IR)\simeq \IR$. The self-dual and
anti-self-dual projections of a two-form $\lambda$ are then given by
$\lambda_{+}=(\lambda,\omega)\omega$,
$\lambda_{-}=\lambda-\lambda_{+}$.
Another important aspect of non-simply connected four-manifolds of $b_2^+=1$
is that
the image of the map
\eqn\map{
\wedge: H^1(X, \IZ) \otimes H^1(X, \IZ) \longrightarrow  H^2(X,\IZ) }
is generated by a single rational cohomology class $\Lambda$
\munoz, so that for any two elements of the basis
$\{\beta_i\}_{i=1, \dots, b_1}$ of $H^1(X,
\IZ)$, $\beta_i \wedge
\beta_j = a_{ij}\Lambda$, where $a_{ij}$ is an antisymmetric
$b_1\times b_1$ matrix \foot{The class $\Lambda$ was denoted by $\Sigma$ in
\mmtwo\munoz.}.

As shown in \mw\mmtwo, the new ingredient in the $u$-plane integral for
non-simply connected manifolds is an integration over the Jacobian torus of
$X$, ${{\bf T} ^{b_1}}=H^1(X, \IR)/H^1(X, \IZ)$. There is a basis of one-forms
on ${\bf T}^{b_1}$ that we will denote by $\beta_i^\sharp
\in H^1 ({\bf T }^{b_1}, \IZ)$, and which are dual to $\beta_i \in  H^1 (X,
\IZ)$.
Notice that there is an isomorphism $H_1(X, \IZ) \simeq H^1({\bf T}^{b_1},
\IZ)$, given
by $\delta_i \rightarrow \beta^{\sharp}_i$. We will then define
$\delta^{\sharp}=
\sum_{i=1}^{b_1} \zeta_i \beta_i^{\sharp}$ as the image of $\delta$ in
\cycles\ under this isomorphism.
Finally, we introduce a symplectic two-form on
${\bf T}^{b_1}$ as
\eqn\volume{
\Omega= \sum_{i<j} a_{ij} \beta^{\sharp} _i \wedge \beta^{\sharp}_j, }
which does not depend on the choice of basis. This is a volume
element for the torus, hence
\eqn\voltee{ {\rm vol}({\bf T}^{b_1})= \int_{{\bf T}^{b_1}}
{\Omega^{b_1/2} \over (b_1/2)!}.}
We can now write the $u$-plane integral in the non-simply connected case:
\eqn\newcpione{
\eqalign{
Z_u &= \left\langle\ex^{p{\cal O}(P)+W_1(\Gamma)+ I(S)}
\right\rangle_{u}^{w_2(E)} \cr
&=-4 \pi i \int_{\gof \backslash \CH}
{dx dy \over  y^{1/2}}
 \int_{{\bf T}^{b_1}}h_{\infty}^{-1}
 \widehat f_{\infty} (p,\delta,S, \tau, y) \Psi(\tilde S),\cr}
}
where $x={\hbox{\rm Re}}(\tau)$, $y={\hbox{\rm Im}}(\tau)$. In this
expression, the
function $\widehat f_{\infty} (p,\delta,S, \tau, y)$ is an almost holomorphic
modular form, as well as a differential form on ${\bf T}^{b_1}$, given by:
\eqn\nwcnvpione{
\widehat f_{\infty} (p,\delta,S, \tau, y)  =
{\sqrt{2} \over  64 \pi }h_{\infty}^{b_1-3}\vartheta_4^{\sigma}
f_{2\infty}^{-1} {\rm e}^{2 p u_{\infty} + S^2 {\widehat T}_{\infty}}
 \exp \biggl[ 2 f_{1 \infty} (S, \Lambda) \Omega + i
h_{\infty}^{-1}\delta^{\sharp} \biggr].}
We have denoted the intersection form in two-cohomology by $(\, , \,)$ and
used Poincar\'e
duality to convert cohomology classes in homology classes.
$\Psi(\tilde S)$ is a Siegel-Narain theta function given by:
\eqn\siegnar{
\eqalign{
\Psi(\tilde S) & = \exp(2\pi i \lambda_0^2)
\exp\bigl[  - { 1 \over  8 \pi y}h_{\infty}^{-2}
\tilde S_-^2 \bigr]\cr &
\cdot \sum_{\lambda\in H^2+ \half w_2(E) }
 \exp\biggl[ - i \pi
\bar\tau (\lambda_+)^2 - i \pi   \tau(\lambda_-  )^2
+ \pi i (\lambda -\lambda_0 ,  w_2(X)) \biggr]  \cr
&
\cdot \exp\bigl[- i h_{\infty}^{-1} (\tilde S_-,\lambda_-) \bigr]  \biggl[
(\lambda_+ ,\omega) +{i \over  4 \pi y} h_{\infty}^{-1} (\tilde S_+,\omega )
   \biggr], \cr}}
where
\eqn\shifts{
\tilde S=  S-16f_{2 \infty} h_{\infty}(\Lambda \otimes \Omega).}
so \siegnar\ is also a differential form on ${\bf T}^{b_1}$, and $2
\lambda_0$ is a
integer lifting of $w_2(E)$ \foot{We have absorbed all the $b_1$ dependent
factors in
the $u$-plane integral of \mmtwo\ in the normalization of the
differential forms on ${\bf T}^{b_1}$, in order to get more compact
expressions.}.
Finally, in the above expressions
$u_{\infty}$, $T_{\infty}$, $h_{\infty}$, $f_{1 \infty}$ and $f_{2 \infty}$
are modular forms defined as follows:
\eqn\qforms{
\eqalign{
u_{\infty}&=\half {{\vartheta_2^4+\vartheta_3^4}\over
(\vartheta_2\vartheta_3)^2}={1\over 8 q^{1/4}}(1+20
q^{1/2}-62q+\cdots),\cr
T_{\infty}&=-{1\over24}\left(  {E_2\over
h^2_{\infty}}-8u_{\infty}\right)= q^{1/4}(1-2 q^{1/2}+6q+\cdots)
,\cr
h_{\infty}(\tau)&= {1
\over 2}
\vartheta_2
\vartheta_3=q^{1/8}(1+2q^{1/2}+q+\cdots),\cr
f_{1\infty}(q)&= {2 E_2+
\vartheta_2^4 +
\vartheta_3^4 \over 3
\vartheta_4^8}= 1+ 24 q^{1/2} + \cdots, \cr
f_{2\infty}(q)&={\vartheta_2
\vartheta_3 \over  2
\vartheta_4^8}= q^{1/8} + 18 q^{5/8}+
\cdots.\cr}
}
In this formulae, $q=\ex^{2\pi i\tau}$, and $\vartheta_i$, $i=2,3,4$ are
the Jacobi theta functions (we follow the notation in \mw).
Notice that $T_{\infty}$ does not transform well under modular
transformations, due to the presence of the second Eisenstein series. In
\nwcnvpione,
we have used the related form
\eqn\rel{
\widehat T_{\infty} = T_{\infty} + { 1\over 8\pi y}h_{\infty}^{-2},}
which is not holomorphic but transforms well under modular transformations.
We also define the related holomorphic function $f_{\infty} (p,\delta, S,
\tau)$
as in \nwcnvpione\ but with $T_{\infty}$ instead of $\widehat
T_{\infty}$.

One immediate application of the $u$-plane integral formalism is
the derivation of the wall-crossing behavior of Donaldson invariants.
As explained in \mw, the integral \newcpione\ has a discontinuous
variation at the cusps of ${\gof \backslash \CH}$ whenever the
cohomology class $\lambda\in H^{2}(X;\IZ)$ is such the period
$\omega\cdot\lambda$ changes sign. We then say that $\lambda$ defines a wall.
The cusps are located at $\tau=i \infty$, $\tau=0$, and $\tau=2$.
The wall-crossing behavior associated to the cusp at infinity
gives the wall-crossing properties of the Donaldson invariants, while
the discontinuous variation of the integral at $\tau=0,2$
must cancel against the contribution to wall-crossing
from the Seiberg-Witten piece $Z_{SW}$. As shown in
\mw, this cancellation completely fixes the
structure of $Z_{SW}$.

The wall-crossing of the $u$-plane integral at $\tau=i \infty$ can be
easily derived by imitating the analysis in section
$4$ of
\mw. The conditions for wall-crossing are $\lambda^2<0$ and
$\lambda_{+}=0$, and one finds \mmtwo:
\eqn\wcross{
\eqalign{
WC(\lambda)
& = -{i\over 2} (-1)^{ (\lambda-\lambda_0,   w_2(X))   }
e^{2\pi i\lambda_0^2} \cr
& \cdot \Biggl[q^{- \lambda^2/2} h_{\infty}(\tau)^{b_1-2}
\vartheta_4^{\sigma} f_{2\infty}^{-1}
\exp\biggl\{ 2p u_{\infty}  +  S^2 T_{\infty}
  -  i  (\lambda , S)/h_{\infty}
\biggr\} \cr
& \cdot  \int_{{\bf T}^{b_1}} \exp \left( 2
f_{1\infty}(q)(S,\Lambda)\Omega +16if_{2\infty}(q) (\lambda,  \Lambda)\Omega
+i {\delta^{\sharp} \over
h_{\infty}} \right)
\Biggr]_{q^0}. \cr}
}
Using the $q$-expansion of the different modular forms, it is
easy to check that the wall-crossing term is different from zero only if
$0>\lambda^2\ge p_1/4$, where $p_1$ is the Pontriagin number of the gauge
bundle
(and $p_1 \equiv w_2(E)^2$ mod $4$).
The expression \wcross\ generalizes the wall-crossing formula of
\gottsche\ to non-simply connected manifolds. Wall-crossing terms
for non-simply connected manifolds were computed in \munoz\ using
algebro-geometric
methods in some particular cases \foot{In comparing to the
expressions in \gottsche\munoz, one has to take into account that what they
call
$\xi$ or $\zeta$ is in fact our $2 \lambda$.}.

\subsec{The Seiberg-Witten contribution}
The structure of the Seiberg-Witten contribution $Z_{SW}$ on
non-simply connected four-manifolds has been studied in detail in
\mmtwo, following the approach in \mw. We will not repeat the
analysis here but we will write down several formulas which will
be useful later.

The SW part $Z_{SW}$ contains two pieces which correspond to the
cusps at $\tau =0,2$, and is written in terms of universal modular
forms and of the Seiberg-Witten invariants introduced in \mfm. A crucial
ingredient in the discussion of the Seiberg-Witten contribution for
non-simply connected manifolds is that we have to consider
generalized Seiberg-Witten invariants, which involve integration of
differential
forms on the moduli space of solutions to the monopole equations. These
differential forms
can be constructed, in the context of topological field theory, using the
descent procedure, and they are associated to one-cycles in the
four-manifold $X$.
Equivalently, to every element $\beta_i$ in the basis of one-forms of $X$
introduced above, with
$i=1, \dots, b_1$, we associate a one-form $\nu_i$ on ${\cal M}_{\lambda}$.
The generalized Seiberg-Witten invariants are then introduced as follows.
Let $\lambda \in H^2(X, \IZ) + w_2 (X)/2$ be a ${\rm Spin}^c$
structure on $X$ \foot{In the mathematical
literature, ${\rm Spin}^c$ structures are rather
given by integral cohomology classes
which reduce to $w_2(X)$ mod 2. They correspond to
$2 \lambda$ in our notation.}, and
let ${\cal M}_{\lambda}$ be the corresponding
Seiberg-Witten moduli space, with virtual
dimension $d_{\lambda}= \lambda^2-(2 \chi + 3 \sigma)/4$.
We then define:
\eqn\gensw{
SW(\lambda, \beta_1\wedge \cdots \wedge \beta_r)=
\int_{{\CM}_{\lambda}} \nu_1
\wedge \cdots
\wedge \nu_r \wedge a_D^{d_{\lambda}-r \over 2},
}
where $a_D$ is a two-form which represents the first Chern class of
the universal line bundle on the moduli space. These generalized invariants
(and
their wall-crossing properties) have been considered in \okonek.

We can now write a general expression for $Z_{SW}$ in the case of a
four-manifold
of $b_2^+=1$. To do this, we first introduce the following modular forms:
\eqn\masforms{
\eqalign{
u_{M}(q_D)&={{\vartheta_3^4+\vartheta_4^4}\over
2(\vartheta_3\vartheta_4)^2}= 1+32q_D+256q_D+\cdots,\cr
h_{M}(q_D)&={1\over2i}\vartheta_3\vartheta_4={1\over2i}(1-4q_D+
4q_D^2+\cdots),
\cr
f_{1M}(q_D)&={{2E_2-\vartheta_3^4-\vartheta_4^4}\over3\vartheta_2^8}=
-{1\over8}(1-6q_D+24q_D^2+\cdots),\cr
f_{2M}(q_D)&={\vartheta_3\vartheta_4\over2i\vartheta_2^8}=
{1\over2^9i}({1\over
q_D}-12+72q_D+\cdots),\cr
T_{M}(q_D)&=-{1\over24}\left ( {E_2\over
h_M^2}-8u_M\right)=\half+8q_D+48q_D^2+
\cdots.\cr
}
}
These modular forms are (up to the
modular weight) the same modular forms as in \qforms\ but evaluated
at $\tau_D=-1/\tau$, that is, $\tau h_{M}(\tau)=h_{\infty}(-1/\tau)$, and so
on. In the above expansions, we have used the dual variable
$q_D={\rm e}^{2 \pi i \tau_D}$. We will denote
by $\delta_{*}=\sum_{i=1}^{b_1} \zeta_i \beta_i$ the formal
combination of one-forms
which is dual to $\delta$ in \cycles. $Z_{SW}$ can then
be written as the sum of two terms.
The first one (corresponding to the cusp at $\tau=0$,
or monopole cusp) is given by the
following expression:
\eqn\swexpand{
\eqalign{
Z_{SW,M}=&\sum_{\lambda} {i ^{b_1 +1} \over 8} \sum_{b\ge0}
\sum_{n=0}^{b} {(-1)^n  \over n! (b-n)!} 2^{-6n-5b+b_1/2}
e^{2i\pi(\lambda_0\cdot\lambda+\lambda_0^2)}\cr
& \cdot \Biggr[ q_D^{-\lambda^2/2} h_M^{b_1-3} \vartheta_2^{8 + \sigma}
\bigl( {i f_{1M} \over
f_{2M}}\bigr) ^{(b+n-b_1)/ 2}  \exp\biggl\{
2pu_M - i  h_M (S, \lambda)
+S^2 T_M(u)\biggr\} \cr
& \,\,\,\,\,\,\,\,\, \biggl( 2 f_{1M} (S, \Lambda) + 16 i \bigl( {1
+ 8 f_{1M}  \over 8 f_{1M}} \bigr)
 f_{2M} (\lambda, \Lambda) \biggr)^n \Biggl]_{q_D^0}\cr
&\cdot \sum_{i_p, j_p=1}^{b_1} a_{i_1 j_1} \cdots a_{i_n j_n}
SW(\lambda, \beta_{i_1} \wedge \beta_{j_1} \wedge \cdots \wedge \beta_{i_n}
\wedge \beta_{j_n} \wedge \delta_{*}^{b-n}),
\cr}
}
where the first sum is over all the ${\rm Spin}^c$-structures
on $X$. As  shown in \mfm,
only a finite number of $\lambda$ give non-zero Seiberg-Witten invariants
(for a given metric).
The second piece in $Z_{SW}$ corresponds to the cusp at $\tau=2$
(the dyon
contribution) and it has exactly the
same form, with the only difference that one has to use the modular forms
\eqn\dyonforms{
u_{dy}=-u_M,\qquad h_{dy}=ih_M,\qquad f_{1dy}=f_{1M},\qquad
f_{2dy}=if_{2M},\qquad T_{dy}=-T_M
}
and include an extra factor $\exp(-2\pi i \lambda^2_0)$. It is easy to
check that
\eqn\dyon{
Z_{SW,dy} (p, \zeta_i, v_i)={\rm e}^{-2\pi i
\lambda^2_0}i^{1-{b_1\over 2}} Z_{SW,M} (-p, i\zeta_i,
-iv_i),}
in agreement with the arguments based on $R$-symmetry \mfm\mw\wittk.
The structure of $Z_{SW}$ is such that the wall-crossing behavior due
to the
Seiberg-Witten invariants exactly cancels the wall-crossing behavior of the
$u$-plane integral at the cusps $\tau=0,2$. For example, at
$\tau=0$ one
can see that \newcpione\ has a discontinuous behavior for
$\lambda \in H^2(X, \IZ) + w_2 (X)/2$
({\it i.e.} a ${\rm Spin}^c$ structure on $X$) such that
$\lambda^2<0$,
and $\lambda\cdot \omega=0$.
These are precisely the conditions for SW wall-crossing. The discontinuity
is given by:
\eqn\swall{
\eqalign{
&{i  \over 8}
e^{2i\pi(\lambda_0\cdot\lambda+\lambda_0^2)} \Biggr[ q_D^{-\lambda^2/2}
h_M^{b_1-3} \vartheta_2^{8 + \sigma}  \exp \biggl\{
2pu_M - i  h_M (S, \lambda)
+S^2 T_M(u)\biggr\} \cr
& \,\,\,\,\,\,\,\,\, \cdot \int_{{\bf T}^{b_1}} \exp \biggl( 2 f_{1M} (S,
\Lambda)\Omega + 16 i
 f_{2M} (\lambda, \Lambda) \Omega + {i \over h_M} \delta^{\sharp}
\biggr)
\Biggl]_{q_D^0}.
\cr}
}
If one compares this expression with \swexpand, one actually recovers
the general wall-crossing formulae of \okonek \foot{Up to a numerical
constant that
appears when one compares the normalization of the fermion fields in the
twisted theory to the
topological normalization of the one-forms $\nu_i$. See \mmtwo\ for details.}:
\eqn\general{
WC(SW(\lambda, \beta_{1} \wedge  \cdots \wedge \beta_{r} ))= { (-1)^{b_1-r\over
2} \over ({b_1-r \over 2})!} (\lambda, \Lambda)^{b_1-r \over 2}
\int_{{\bf T}^{b_1}} \beta^{\sharp}_{1} \wedge  \cdots \wedge
\beta^{\sharp}_{r}\wedge \Omega^{b_1-r\over 2}.}

\newsec{Donaldson invariants of product ruled surfaces}

In this section we will derive explicit results for the Donaldson
invariants on  product ruled surfaces, that is, on four-manifolds of
the form $X=\Sigma_g\times\bs2$, where $\Sigma_g$ is a Riemann
surface of genus
$g$. For these surfaces, $b_1=2g$, $b_2=2$, $b_2^+=1$, so $\sigma=0$
and
$\chi=4-2b_1$. $H^2(X, \IZ)$ is generated by the cohomology classes
$[\bs2]$, $[\Sigma_g]$, with intersection form
\eqn\intform{ II^{1,1}=\pmatrix{0&1\cr 1&0\cr}.}
These surfaces are Spin, therefore $w_2(X)=0$.
The basis of one forms on $\Sigma_g\times\bs2$ is given by the duals
to the usual symplectic basis of one cycles on $\Sigma_g$, $\delta_{i}$,
$i=1, \dots, 2g$,
with $\delta_i \cap \delta_{i+g} =1$, $i=1, \dots, g$. The matrix $a_{ij}$
is then the
symplectic matrix $J$, and
$\Lambda=[\bs2]$ (the Poincar\'e dual to the two-homology class
of
$\bs2$). It follows that
\eqn\volelerul{
\Omega =\sum_{i=1}^g \beta_i^\sharp \wedge \beta_{i+g}^{\sharp}.}
and ${\rm vol}({\bf T}^{b_1})=1$. As it will become clear in the computation,
all the Donaldson polynomials involving the cohomology classes
associated to one-cycles can be expressed in terms of the
${\rm Sp}(2g, \IZ)$-invariant element in $\wedge^{\rm even} H_1 (X, \IZ)$
given by $\iota =\sum_{i=1}^g \delta_i \delta_{i+g}$. This element
of ${\bf A}(X)$ corresponds to the
degree $6$ differential form on the moduli space of instantons given by:
\eqn\invform{
\gamma=-2 \sum_{i=1}^g I(\delta_i)I(\delta_{i+g}).}
If we write $S=s\Sigma_g + t{\bf S}^2$, we see that the generating function
that we
want to compute is
$Z_{DW}(p, r,s,t)={\cal D}_X^{w_2(E)}({\rm e}^{p x + r\iota + s\Sigma_g +
t{\bf S}^2})$.
In the previous section we have computed the generating functional
including $\delta$.
If we want to include  $\iota$ in the $u$-plane integral, we just take into
account
that the  $3$-class $I(\delta_i)$ on the moduli space gives $(i
/h_{\infty}) \beta_i^{\sharp}$
in the $u$-plane integral. Therefore, using \volelerul, we find the
correspondence
\eqn\change{
\gamma \rightarrow {2r \over h_{\infty}^2} \Omega,}
and to obtain $Z_{DW}(p, r,s,t)$ from the above formulae we just have
to change $(i /h_{\infty}) \delta^{\sharp}$ by \change\ in the $u$-plane
integral.
For the Seiberg-Witten contribution, the modification is very similar.

As $b_2^+=1$,  the
generating function (or Donaldson series) for the Donaldson
invariants will  be given by the sum of the $u$-plane integral and
the SW contributions. The  resulting Donaldson polynomials are not topological
invariants. In this case, it is interesting to compute the polynomials in
limiting
chambers, {\it i.e.}. in chambers where one of the factors in the
product is very small (and the other factor is then very big). Once the
invariants
are known in these chambers, we can compute the invariants in any other
chamber by adding a sum of wall-crossings. But the most important reason to
study
the invariants in the limiting chambers is that, for special choices
of the second Stiefel-Whitney classes, the Donaldson polynomials have a
simple structure.
Moreover, the connection to Fukaya-Floer theory involves the limiting chamber
in which $\Sigma_g$ is small.

The limiting chambers can be analyzed in a fairly simple way using the
general expression of the period point, as in \mmtwo:
\eqn\period{
\omega (\theta) ={1 \over {\sqrt 2}}({\rm e}^{\theta}[{\bf S}^2]+ {\rm
e}^{-\theta}[\Sigma_g] ). }
The limiting chambers are $\theta \rightarrow \pm \infty$, which correspond
to the limit
of small volume for $\bs2$ and $\Sigma_g$, respectively. As
explained in \mmtwo, in the
chamber where ${\bf S}^2$ is small and $\theta \rightarrow \infty$, the
scalar curvature
is positive and the Seiberg-Witten invariants vanish \mfm. This has two
important consequences:
first, in this chamber the Donaldson invariants are given just by the
$u$-plane integral. Second,
the Seiberg-Witten invariants in any other chamber
can be computed by wall-crossing and they will be given by the topological
expression \general.
In particular, the SW contribution to the Donaldson invariants will be
given by wall-crossing
of the $u$-plane integral at the cusps $\tau=0, 2$, and we can use the
simple expression
\swall.

\subsec{Computing the $u$-plane integral}
We start by computing the $u$-plane integral. We follow closely the analysis
in section $8$ of \mw.  We first rewrite \newcpione\ as
\eqn\calg{
{\cal G}(\rho)=\int_{\Gamma^{0}(4)\setminus {\cal H}} {dxdy\over y^{3/2}}
\int_{{\bf T}^{b_1}}\hat f_{\infty}(p,S,\tau,y)\bar\Theta,
}
where $\hat f_{\infty}$ is given in \nwcnvpione\ and $\bar\Theta$ is
a Siegel-Narain theta function
introduced in \mw:
\eqn\teta{\eqalign{
\bar\Theta&=
\exp\bigl[  { 1 \over 2y}(\bar\xi_{+}^2-\bar\xi_{-}^2)\bigr]\cr &
\sum_{\lambda\in H^2+ \beta}
\exp\biggl[ - i \pi
\bar\tau (\lambda_{+})^2 - i \pi   \tau(\lambda_{-}  )^2
- 2\pi i (\bar\xi,\lambda)+2\pi i(\lambda,\alpha) \biggr]
,\cr
}}
with
\eqn\ji{
\bar\xi=\rho y h_{\infty}\omega+{1\over 2\pi h_{\infty}}\tilde S_{-}
,\qquad \alpha=0,\qquad \beta=\half w_2(E).}
$Z_u$ is obtained from ${\cal G}$ by
\eqn\calgi{
Z_u=(\tilde S,\omega){\cal G}(\rho)\vert_{\rho=0}+2{d{\cal G}\over d\rho}
\vert_{\rho=0}.
}
Next we
bring the integral \newcpione\
over $\Gamma^{0}(4)\setminus {\cal H}$ to an
integral
over the fundamental domain of $SL(2,\IZ)$. Recall that a fundamental domain
for $\Gamma^{0}(4)$ can be obtained from a fundamental domain
${\cal F}$ for
$SL(2,\IZ)$ as follows
\eqn\fundom{
\Gamma^{0}(4)\setminus {\cal H}\cong {\cal F}\cup (T\cdot{\cal F})\cup
(T^2\cdot{\cal F})\cup (T^3\cdot
{\cal F})\cup (S\cdot{\cal F})\cup (T^2S\cdot{\cal F}),
}
where $T$ and $S$ are the standard generators of  $SL(2,\IZ)$.
The first four domains correspond to the cusp at $\tau\to i\infty$ and will
be referred to as the semiclassical cusp. The domain $S\cdot{\cal F}$
corresponds to the cusp at $\tau=0$  (the monopole
cusp.) The domain $T^{2}S\cdot{\cal F}$ corresponds to the cusp at
$\tau=2$,        or dyon cusp.

Taking this into account, we can bring the integral \calg\ to the form
\eqn\calgii{
{\cal G}(\rho)=\int_{\Gamma^{0}(4)\setminus {\cal H}} {dxdy\over
y^{2}}\int_{{\bf T}^{b_1}} \hat f y^{1/2}\bar\Theta=
\int_{\Gamma^{0}(4)\setminus {\cal H}} {dxdy\over y^{2}}\int_{{\bf
T}^{b_1}}  {\cal I}(\tau)= \int_{{\cal F}} {dxdy\over
y^{2}}\sum_{I}\int_{{\bf T}^{b_1}}{\cal I}_{I}(\tau), }
where ${\cal
I}_I=\hat f_I\, y^{1/2}\,\bar\Theta_I$ denotes the modular forms
\eqn\cali{\eqalign{
{\cal I}_{(\infty,0)}(\tau)&= {\cal I}(\tau),\cr
{\cal I}_{(\infty,1)}(\tau)&= {\cal I}(\tau+1),\cr
{\cal I}_{(\infty,2)}(\tau)&= {\cal I}(\tau+2),\cr}
\qquad\qquad
\eqalign{
{\cal I}_{(\infty,3)}(\tau)&= {\cal I}(\tau+3),\cr
{\cal I}_{M}(\tau)&= {\cal I}(-1/\tau),\cr
{\cal I}_{dy}(\tau)&= {\cal I}(2-1/\tau).\cr
}}

Therefore, the integral splits into $6$ integrals over the modular domain of
$SL(2,\IZ)$. Now we can analyze each of these following the general procedure
described in section $8$ of \mw. From the modular
properties of the theta function \teta\ -- see for example appendix B of \mw\
-- one notes that the corresponding theta functions
have
\eqn\fluxes{\matrix{
&(\infty,0)&(\infty,1)&(\infty,2)&(\infty,3)&M&dy\cr\cr
\alpha^I&0&w_2(E)/2&0&w_2(E)/2&w_2(E)/2&w_2(E)/2\cr\cr
\beta^I&w_2(E)/2& w_2(E)/2&w_2(E)/2&w_2(E)/2&0&0&\cr}
}
where we have taken into account that $w_2(X)=0$ for product ruled
surfaces.
The computation of the $u$-plane integral
proceeds as in \mw, following a general strategy due to
Borcherds \borchaut, which is a generalization of standard
techniques in one-loop threshold corrections in string theory --
see for example
\harvmoore. This strategy is called lattice reduction.
To perform the
lattice reduction, one has to choose a
reduction vector $z$ in the cohomology lattice.
$z$ must be a primitive vector of zero norm,
and the computation using lattice
reduction will be valid in the chamber
with $(z,\omega)^2$ very small. We have then
two possible choices of $z$ depending on
the limiting chamber we choose: for small
$\Sigma_g$, one has $z=[\Sigma_g]$, and for
small ${\bf S}^2$ one has $z=[{\bf S}^2]$.
One also chooses another norm zero vector $z'$
such that $(z,z')=1$. The value of the $u$-plane
integral depends of course on the value of the
Stiefel-Whitney class, that we can write
as \eqn\flujo{
w_2(E)=\epsilon z + \epsilon ' z',}
where $\epsilon, \epsilon' =0,1$.
We then write
\eqn\labe{
\beta^I= {q^I \over 2} z + {r^I \over 2}z'}
where $q^I=\epsilon$, $r^I=\epsilon '$ for the cusps at infinity, and
are zero otherwise. We write the elements in the lattice $H^2(X, \IZ)$ as
$cz' + nz$,
where $c, n \in \IZ$.
The contribution of the $I$-th cusp to the integral \calg\ reads,
after a Poisson summation on $n$,
\eqn\cdint{
\eqalign{
&{\ex^{2\pi i\lambda_0^2}\over
\sqrt{2  z_+^2}} \int_{\CF }
{dx dy \over  y^{2}} \int_{{\bf T}^{b_1}}
\hat f_{I}
\sum_{c, d\in \IZ}
\exp\biggl[-{\pi \over  2 y z_+^2}\vert (c+ {r^I / 2})  \tau
+ d \vert^2-{\pi \over  y z_+^2} (\bar \xi_+^I, z_+)(\bar
\xi_-^I,z_-) \biggr]
\cr
&\exp\Biggl[ - {\pi (\bar\xi_+^I, z_+) \over    z_+^2}
\Bigl({ (c+ r^I / 2)   \tau + d \over  y}\Bigr)
- {\pi (\bar \xi_-^I, z_-) \over    z_+^2}
\Bigl({ (c+ r^I / 2)   \bar \tau + d \over  y} \Bigr)+{\pi \over  y z_+^2}
(\bar \xi^I, z)(\alpha^I,z)\Biggr]\cr
&\exp\Biggl[-i \pi q^I d -{\pi \over 2 y z_+^2}(\alpha^I,z)^2+ \pi
{(\alpha_+^I,z_+)\over y z_+^2} \Bigl((c+ {r^I\over 2})\bar\tau+d \Bigr) + \pi
{(\alpha_-^I,z_-)\over y z_+^2}\Bigl((c+ {r^I \over 2})\tau+d \Bigr)
\Biggr].
\cr
}
}
We will now analyze the $u$-plane integral in the two limiting chambers.
We first consider
the case of ${\bf S}^2$ small. In this case, $z=[{\bf S}^2]$. The first
thing to notice is
that, if $\epsilon' \not= 0$, then the cusps at infinity do not give any
contribution.
This is due to the non-zero $r^I$ in the exponentials in \cdint, and it can
be proved using
the analysis in section 5 of \mw. Moreover, the cusps at $\tau=0, 2$ do not
contribute
either, because the measure in the $u$-plane integral goes like $q_D+
\dots$. Therefore,
if $(w_2(E),[{\bf S}^2])\not =0$, the $u$-plane integral vanishes in the
chamber where
the volume of $[{\bf S}^2]$ is very small. This is an example of the
vanishing theorem of \mw.

We then have to consider only the case of $w_2(E) = \epsilon [{\bf S}^2]$.
In this
case, the contribution comes from the cusps at infinity, which have $q^I
=\epsilon$.
Notice that $\alpha^I = (\epsilon/2) z$ for $I=(\infty,1)$ and
$(\infty,3)$. It is easy to
see that the inclusion of $\alpha^I$ for these cusps is equivalent to an
extra phase
$-\pi i \epsilon c$. Following \mw, we now apply the unfolding technique to
the integral \cdint. The action of $SL(2,\IZ)$ on
$c$ and $d$ has two classes of orbits:  non-degenerate orbits with
$c$, $d$ not both zero, and the degenerate orbit with $c=d=0$.
Non-degenerate orbits can be transformed by
$SL(2,\IZ)$ to have $c=0$,
giving an integral over a strip $0\leq x\leq 1$ in the
upper half plane, together with
a sum over $d\in \IZ\setminus\{0\}$. In this case, the contribution
of the degenerate orbit can be combined with the contribution of the
non-degenerate orbits. As $c=0$ in any case,
the four cusps at infinity give the same contribution and they add
to
\eqn\nondegii{
- 8 \sqrt{2}\ex^{2\pi i\lambda_0^2}
\Biggl[\int_{{\bf T}^{b_1}}
f_{\infty} h_\infty
\sum_{ d=-\infty}^{\infty}
{{\rm e}^{-\pi i \epsilon d }
\over
d +A_{\infty} (z)  }
 \Biggr]_{q^0},
}
where
\eqn\laa{
A_I (z)\equiv {(\widetilde S, z)\over 2 \pi h_{I}}.
}
In this case, as $z=[{\bf S}^2]$ and therefore $(\Lambda,z)=0$, one
has
\eqn\laaaqui{
A_{\infty}(z)={s \over 2\pi h_\infty}.}
Using now the identity,
\eqn\triide{
\sum_{ d=-\infty}^\infty {{\rm e}^{i\theta d} \over  d+B}=
{2 \pi i } {{\rm e}^{-iB\theta} \over  1-e^{-2 \pi i B}
},
}
which is valid for $0\le \theta <2 \pi$,
and integrating over ${\bf T}^b_1$, one finally obtains the expressions
\mmtwo
\eqn\ruledexp{
Z^{w_2(E)=0}_{g, {\bf S}^2}=-{i\over 4}
\biggl[ (h_{\infty}^2 f_{2\infty})^{-1} {\rm e} ^{2pu_{\infty} +
S^2T_{\infty}}
\Bigl( 2 f_{1\infty} h_{\infty}^2 s+2r\Bigr)^g
\coth \Bigl( {i s \over 2h_{\infty}} \Bigr)
\biggr]_{q^0}, }
where we used \triide\ with $\epsilon=0$, as in the original computation
in \mw, and
\eqn\ruledexpf{
Z^{w_2(E)=[{\bf S}^2]}_{g, {\bf S}^2}=-{1\over 4}
\biggl[ (h_{\infty}^2 f_{2\infty})^{-1} {\rm e} ^{2pu_{\infty} +
S^2T_{\infty}}
\Bigl( 2 f_{1\infty} h_{\infty}^2 s+2r\Bigr)^g
\csc \Bigl( { s \over 2h_{\infty}} \Bigr)
\biggr]_{q^0},}
where we used again the identity \triide\ with $\theta=\pi$. This changes
the $\coth$ in a $\csc$. For $g=0$, one recovers the expressions
for ${\bf S}^2 \times {\bf S}^2$ which were obtained in \mw\gottzag.

Let us now consider the other limiting chamber, in which the volume of
$\Sigma_g$ is
small. We then have $z=[\Sigma_g]$. We will also
restrict ourselves to the Stiefel-Whitney classes of the
form $w_2(E)=[{\bf S}^2] + \epsilon [\Sigma_g]$, {\it i.e.},
$\epsilon ' =1$. The reason for this is simple: in this case, $r^I=1$ for
the cusps at infinity and the vanishing argument of \mw\ applies.
Therefore, there
is no contribution from these cusps. In other words,
$(\beta^I,[\Sigma_g])\not=0$ and
there is a non-zero magnetic flux through the vanishing fiber
$\Sigma_g$. Notice that this does not imply that the whole $u$-plane integral
is zero. As it has been already remarked at the end of section 5 in \mw,
the vanishing
of the monopole and dyon cusps depends crucially on the behavior of the
measure.

Let us then focus on the contribution from the monopole cusp (the
contribution from the dyon cusp is related to that of the monopole
cusp by \dyonforms). It is easy to see that we can absorb the
$\alpha^M$ dependence in \cdint\ in a shift $d\to d-\half$ plus an
extra
$\tau$-independent phase $-i\pi\epsilon c$. We can now apply the unfolding
procedure. The shift in $d$ is crucial to the final result. There is a subtlety
here, since $c$ and $d$ come in the combination
$c\tau+\tilde d$, where $\tilde d=d-\half$. $SL(2,\IZ)$ elements
$\pmatrix{\alpha&\beta\cr
\gamma&\delta\cr}$ with $\gamma$ odd make
$c$ half-integer (and hence not zero) and those with $\delta$
even result in $\tilde d\in \IZ$.  Both problems are related and can
be solved as follows. If $\gamma$ is odd, $c'=\alpha c+\gamma
(d-\half)\in\IZ-\half$. Looking back to \cdint, we see that a
half-integer $c$ corresponds to having a non-zero flux through
the vanishing fiber, so the integral corresponding to those values
of $c$ vanishes. If $\delta$ is even, then, as $\alpha\delta-
\beta\gamma=1$ (which means in particular that $\gamma$ and
$\delta$ are coprime), $\gamma$ is necessarily odd, and the
integral vanishes as well. So we do not get any contribution from
orbits with $c\in \IZ+\half$ or $\tilde d\in \IZ$. Therefore, the final
result is the same as in equation (C.2) in \mw\ (with minor changes):
\eqn\nondegmon{
- 2 \sqrt{2}\ex^{2\pi i\lambda_0^2}
\Biggl[\int_{{\bf T}^{b_1}}
f_{M} h_M
\sum_{ d=-\infty}^{\infty}
{1
\over
d -{1 \over 2} +A_M (z)  }
 \Biggr]_{q_{D}^0},
}
where
\eqn\laa{
A_M (z)= {t \over 2 \pi h_{M}}-{8 \over \pi}f_{2M}\Omega.
}
and $f_M$ is the magnetic version of \nwcnvpione:
\eqn\efemon{
f_{M}h_{M}={{\sqrt 2} \over 64} h_{M}^{2(g-1)}f_{2M}^{-1}
{\rm e}^{2 p u_{M} + S^2 T_{M}}\exp
\biggl[ 2\Big(f_{1M}s +{r \over h_{M}^2}\Big)\Omega
\biggr].
}
Using the identity \triide\ with $B=A_M(z) -1/2$, we finally obtain the
final result for the $u$-plane integral in the limiting chamber
where the volume of $\Sigma_g$ is small:
\eqn\uplane{\eqalign{Z_{u, \Sigma_g}^{w_2(E)}&=-4\raiz\pi i \ex^{2\pi
i\lambda_0^2}
\Bigg\{\Big[\int_{{\bf T}^{b_1}}{f_M h_M\over
1+\ex^{-2\pi iA_M}}
\Big]_{q_{D}^0}+\ex^{-2\pi i\lambda_0^2}\Big[\int_{{\bf
T}^{b_1}}{f_{dy} h_{dy}\over 1+
\ex^{-2\pi iA_{dy}}}\Big]_{q_{dy}^0}\Bigg\}\cr
}.}
To obtain an explicit expression for \uplane, we have to integrate over
the Jacobian ${\bf T}^{b_1}$. To do this, we have to expand the exponential
involving
$\Omega$ in the numerators of \uplane. Notice that
\eqn\expan{
{1 \over 1 + {\rm e}^{t+x}} = {1 \over 1 +{\rm e}^{t}} +
 \sum_{m\ge 1}{\rm Li}_{-m} (-{\rm e}^t)
{ x^m \over m!},}
where ${\rm Li}_n$ is the polylogarithm of index $n$. We have taken into
account
the fact that, for negative index, the polylogarithm is given by
\eqn\neg{
{\rm Li}_{-m}({\rm e}^t) = \bigl( {d \over dt} \bigr) ^{|m|} {1 \over
1-{\rm e}^t},}
and using this, the identity \expan\ follows immediately. One also has:
\eqn\polyone{
{\rm Li}_{-m}(-1)={1 \over 2} E_m = -{ 1\over m+1} (2^{m+1}-1) B_{m+1},}
where $E_m$ are the Euler numbers and $B_m$ are the Bernoulli numbers.
To write the final expression for the
$u$-plane integral, we also define the normalized modular forms:
\eqn\aunmasforms{
\tilde h_{M}=2ih_M,\qquad \tilde f_{1M}=-8f_{1M},\qquad
\tilde f_{2M}= 2^9 i f_{2M}.
}
The $u$-plane integral
for the Stiefel-Whitney class $w_2(E)=[\bs2]+\epsilon[\Sigma_g]$,
$\epsilon =0,1$,
is then given by:
\eqn\uplanetwo{
\eqalign{Z_{u,\Sigma_g}^\epsilon(p,r,s,t)
=&-2^8 (-1)^\epsilon
\Bigg[ {\rm e}^{2pu_M+2st T_M}
\sum_{m=1}^g {g \choose m} (-1)^m 2^{-6m}(\tilde h_M^2\tilde f_{2M})^{m-1}\cr
&\,\,\,\,\,\,\  \quad \quad \quad \cdot
\biggl( 2r + {s \over 16} \tilde h_M^2 \tilde f_{1M} \biggr)^{g-m} {\rm
Li}_{-m}
(-{\rm e}^{2t\tilde h_M^{-1}})\Bigg]_{q_D^0} \cr
& -2^8 i^{1-g}
\Bigg[ {\rm e}^{-2pu_M-2st T_M}
\sum_{m=1}^g {g \choose m}(-1)^m 2^{-6m}(\tilde h_M^2\tilde f_{2M})^{m-1}\cr
&\,\,\,\,\,\,\  \quad \quad \quad \cdot
\biggl( 2ir - {is \over 16} \tilde h_M^2 \tilde f_{1M} \biggr)^{g-m} {\rm
Li}_{-m}
(-{\rm e}^{-2it\tilde h_M^{-1}})\Bigg]_{q_D^0} \cr
}}
where we have used that
$\lambda_0^2=w_2(E)^2/4~(\mod~\IZ)=\epsilon/2~(\mod~\IZ)$. The first
piece corresponds to the monopole cusp at $\tau=0$, while the second one
corresponds
to the dyon cusp at $\tau=2$. Notice that in the above expression we have not
included the term $m=0$ in the sum which comes from the expansion \expan.
This is due to the
fact that this term has an overall $\tilde f_{2M}^{-1} =q_D + \cdots$. As
the rest of the
modular forms are analytic in $q_D$, there can not be any $q_D^0$ term in
the expansion, so this
term does not contribute.

\subsec{The Seiberg-Witten contribution}

Let us now turn to the Seiberg-Witten contribution. As we have already
remarked, the
Seiberg-Witten invariants vanish in the chamber
of small volume for ${\bf S}^2$, and the Seiberg-Witten invariants in the
chamber where $\vol(\Sigma_g)\to 0$ are given by the sum of
wall-crossing terms of the form \general. But these terms match the
wall-crossings
at the cusps $\tau=0,2$ of the $u$-plane integral. Therefore, the SW
contribution
to the Donaldson invariants is given by the sum of wall-crossings \swall.
Which walls do we cross in going from the limiting chamber where
$\vol({\bf S}^2)\to 0$ to the chamber where  $\vol(\Sigma_g)\to 0$?
Any $\lambda$ with $\lambda^2<0$ and $d_{\lambda}=
\lambda^2-(2\chi+3\sigma)/4=\lambda^2+2(g-1)\geq 0$ defines a wall. If we
set $\lambda=-b[\Sigma_g]+a[{\bf S}^2]$, $a,b\in\IZ$, these
conditions give:
\eqn\swwccond{
0<ab\leq g-1.}
We then see that, for a fixed genus $g$, there is only a finite number of walls
for the Seiberg-Witten contribution. This is in sharp contrast with the
wall-crossing
terms coming from the cusps at infinity, analyzed in \mmtwo, where there is
an infinite number
of walls. Notice that $\lambda$ and $-\lambda$ define the same wall, but
the wall-crossing
term has opposite sign. We then obtain the following expression for the
Seiberg-Witten
contribution:
\eqn\switten{\eqalign{
Z_{SW, \Sigma_g}^{\epsilon}(p,r,s,t)=
& 2^{8}(-1)^{\epsilon}
\sum_{\scriptstyle 0<ab\leq g-1} \Bigg[
{\rm sgn (a)} q_D^{ab}(-1)^{a\epsilon+b}\,(\tilde
h_M^2\tilde f_{2M})^{-1}\,\ex^{2p u_M+ 2 s t T_M} \cr
&\cdot \Biggl(
\biggl( {s\over 16 } \tilde h_M^2 \tilde f_{1M}
+2^{-7} b \tilde h_M^2\tilde f_{2M}+ 2r \biggr)^g \ex^{{2as-2bt\over \tilde
h_M}} \Biggr)
\Bigg]_{q_D^0}\cr
& +2^{8}i^{1-g}
\sum_{\scriptstyle 0<ab\leq g-1} \Bigg[{\rm sgn (a)}
q_D^{ab}(-1)^{a\epsilon+b}\,(\tilde h_M^2\tilde
f_{2M})^{-1}\,\ex^{-2pu_M-2 s t T_M} \cr
&\cdot \Biggl(
\biggl( -{i s\over 16 }\tilde h_M^2 \tilde f_{1M}
+ 2^{-7} b \tilde h_M^2 \tilde f_{2M}+ 2i r \biggr)^g
\ex^{-{2ias-2i b t\over \tilde h_M}} \Biggl)
\Bigg]_{q_D^0},\cr }}
where again the first piece corresponds to the monopole cusp, and the
second piece to the dyon cusp. The ${\rm sgn}(a)$ appears because, as
we explained before, the wall-crossing terms corresponding to $\lambda$ and
$-\lambda$
have opposite sign. The overall sign can be fixed by looking at
the sign of the Seiberg-Witten invariants in the wall-crossing
formula of \okonek. Notice that the constraint $g-1\ge ab$ is in fact
redundant, as the whole expression above vanishes if this constraint is not
fulfilled. This is due to the fact that the most negative power of $q_D$ in the
above expression comes from $\tilde f_{2M}^{g-1} =q_D^{1-g}(1 + {\cal
O}(q_D))$.
Therefore, if $ab>g-1$, there are no terms in $q_D^0$.

The Donaldson-Witten generating
function in
the chamber where $\Sigma_g$ is small, and for $w_2(E)=[{\bf S}^2]+\epsilon
[\Sigma_g]$, is then given by the sum of \switten\ and
\uplanetwo, and we will denote it by $Z^{\epsilon}_{\Sigma_g}(p,r,s,t)$.
Notice that $Z^{\epsilon}_{\Sigma_g}(p,r,s,t)$ can be also computed by adding
the generating function in the chamber of small volume for ${\bf S}^2$ and
the infinite sum of wall-crossings. The resulting expressions were given in
\mmtwo, in terms of Weierstrass $\sigma$ functions. The fact that
the generating functions in \mmtwo\ are equal to
$Z^{\epsilon}_{\Sigma_g}(p,r,s,t)$
gives a remarkable identity. They encode the information about the
Donaldson invariants in two different ways, that we can call ``magnetic" and
``electric." We have checked that they give indeed the same invariants in
many cases,
but an analytic proof would be rather formidable.

It is interesting to notice that the $u$-plane integral can be
written as
\eqn\uplane{
\eqalign{
-2^8 (-1)^\epsilon
&\Bigg[{\rm e}^{2pu_M+2st T_M}(\tilde h_M^2\tilde f_{2M})^{-1}\cr
& \,\,\,\,\,\,\ \cdot \biggl( 2r + { s \over 16}  \tilde h_M^2 \tilde
f_{1M} -2^{-8}
\tilde f_{2M}\tilde h_M^3 {d \over d t} \biggr)^g
\sum_{n=0}^{\infty} (-1)^n {\rm e}^{{2n t \over \tilde h_M}}\Bigg]_{q_D^0},
\cr}}
where we have only written the monopole contribution. This piece has the
same structure of the Seiberg-Witten contribution \switten, but where the sum
is now over an {\it infinite} number of basic classes of the form $\lambda =
n[\Sigma_g]$, {\it i.e.} with $b=-n\le 0$ and $a=0$. Similar considerations
have been made in \gottzag\ for simply-connected manifolds.

\subsec{Some properties and examples}

An interesting corollary of our computation is that the manifold
$\Sigma_g \times {\bf S}^2$ is of $g^{th}$ finite type, when one considers
the chamber of small volume for $\Sigma_g$ and a Stiefel-Whitney class such
that
$(w_2(E), \Sigma_g)=0$. This is an immediate consequence of our expressions.
To see it, notice that both for the $u$-plane and the SW contributions, the
only source of
a negative power of $q_D$ is the term $\tilde f_{2M}$. The rest of the
modular forms
involved in our formulae are analytic in $q_D$. On the other hand, the
maximum possible
power of $\tilde f_{2M}$ is precisely $g-1$. As $u_M =1 + 32 q_D + \dots$,
an insertion of
$(u_M \pm 1)^g$ in the monopole (respectively, dyon) contribution to
\uplane\ or \switten,
will make the generating
function vanish. But an insertion of $u_M \pm 1$ is equivalent to acting with
\eqn\opact{
{\partial \over \partial p} \pm 2 }
on the Donaldson-Witten generating function. We then find that
\eqn\finite{
\biggl( {\partial^2 \over \partial p^2} - 4 \biggr)^g
Z_g^{\epsilon}(p,r,s,t)=0.}
$g$ is in fact the minimum power we need to kill
the generating function, since
\eqn\menosun{\eqalign{
&\biggl( {\partial^2 \over \partial p^2} - 4 \biggr)^{g-1}
Z_g^{\epsilon}(p,r,s,t)= \cr
&\quad \quad (-1)^{\epsilon+g-1}2^g \Bigl( {\rm Li}_{-g}(-{\rm e}^{2t}){\rm
e}^{2p + st} +
(-1)^{\epsilon}i^{1-g} {\rm Li}_{-g} (-{\rm e}^{-2it}) {\rm e}^{-2p
-st}\Bigr).\cr}}
We conclude that $\Sigma_g \times {\bf S}^2$ is of $g^{th}$
 finite type for the chamber and the Stiefel-Whitney class under
consideration. This
was proved in \msz\ for $g=1$.

We will now give explicit expressions for the Donaldson-Witten generating
function
at low genus. For $g=1$, the Seiberg-Witten contribution vanishes as there
are no walls
({\it i.e.}, the conditions \swwccond\ have no solution). In this case,
only the
$u$-plane contributes. The only polylogarithm involved here is
\eqn\polipart{
{\rm Li}_{-1}(-{\rm e}^t)= -{{\rm e}^t \over (1+{\rm e}^t)^2}=-{1 \over
4(\cosh (t/2))^2}.}
It is clear that for $g=1$ we only need the first term in the expansion of the
modular forms. We then find,
\eqn\moszaboi{
Z_{1, \Sigma_1}^{\epsilon}(p,r,s,t)=Z_{u,
\Sigma_1}^{\epsilon}(p,r,s,t)=-\half (-1)^{\epsilon}\left [
{\ex^{2p+st}\over \cosh^2 (t)}+(-1)^\epsilon
{\ex^{-2p-st}\over \cosh^2(-i t)}\right].
}
As in this case the manifold has a simple type behavior, we can define the
Donaldson series $\ID^{\epsilon}=Z_{DW}^{\epsilon}
\vert_{p=0}+\half {\partial\over\partial p}Z_{DW}^{\epsilon}\vert_{p=0}$. If we
write it as a functional on ${\rm Sym}(H^2(X, \IZ))$, we find from \moszaboi:
\eqn\moszaboii{
\ID^{\epsilon}=(-1)^{(e^2-2e\cdot F) }
{\ex^{Q/2}\over \cosh^2 F },}
where $e=w_2(E)$, $F=[\Sigma_1]$, and $Q$ is the intersection form. This
is in perfect agreement with the Theorem 1.3 in \msz.

It is clear that, as we consider larger values of $g$, the expression for
the Donaldson-Witten
generating function becomes more and more complicated. The main source of this
complexity is the $t$-dependence in the $u$-plane integral. For example,
for $g=2$
the monopole contribution to the $u$-plane integral is given by:
\eqn\genustwo{
\eqalign{
Z_{u,M}^{\epsilon}=&-(-1)^{\epsilon}{ {\rm e}^{2p +st + 2t} \over 16 (1 +
{\rm e}^{2t})^4}
  \bigl( 5 - 5 {\rm e}^{4t} + 16 \left( -1 + {\rm e}^{4 t} \right)  p +
    128  \left( 1 + {\rm e}^{2 t} \right) ^2 r + 4s \cr
& \,\,\,\,\,\,\,\,\,\,\,\,\,\,\,\,\,\, + 8 {\rm e}^{2t} s +
    4{\rm e}^{4 t}s - 2t + 8{\rm e}^{2t} t -
    2{\rm e}^{4t} t - 4s t +
    4{\rm e}^{4 t}s t  \bigr) , \cr}
    }
while the Seiberg-Witten contribution is
\eqn\swgenus{
Z_{SW,M}^{\epsilon}=(-1)^{\epsilon}{{\rm e}^{2p + st} \over 64}
 \bigl( {\rm e}^{2t-2s} - {\rm e}^{2s-2t} \bigr).}

\newsec{Application 1: intersection pairings on the moduli
space of stable bundles}

\subsec{The moduli space of stable bundles}
Let $\Sigma_g$ be a Riemann surface of genus $g$. The moduli space of
flat $SO(3)$ connections on $\Sigma_g$, with Stiefel-Whitney class
$w_2\not= 0$ turns out to be a very rich and interesting space.
One of
the reasons for this richness is the fact that this moduli space can be
understood
in many different ways: using the Hitchin-Kobayashi correspondence, we can
think
about this space as the moduli space of rank two, odd degree stable bundles
over
$\Sigma_g$ with fixed determinant. On the other hand, due to the classical
theorem
of Narashiman and Seshadri, we can identify this moduli space with the
representations in
$SU(2)$ of the fundamental group of the punctured Riemann
surface $\Sigma_g \backslash D_p$, where $D_p$ is a
small disk around the puncture $p$, and with holonomy $-1$
around $p$ (the fact that we require a non-trivial holonomy is due precisely
to the non-zero
Stiefel-Whitney class). In any case, this moduli space, that we will denote by
${\cal M}_g$, is
a smooth projective variety of (real) dimension $6g-6$. Similarly, we can
consider the moduli space
of flat $SU(2)$ connections, {\it i.e.} with $w_2=0$. This moduli space can
be identified
with the moduli space of stable rank two vector bundles of {\it even}
degree and it is singular.
We will denote it by ${\cal M}_g^+$.

The cohomology ring of ${\cal M}_g$ can be studied by using
a two-dimensional version of the
$\mu$ map which arises in Donaldson theory. This map sends homology classes
of $\Sigma_g$
to cohomology classes of ${\cal M}_g$. The generators of $H_* (\Sigma_g)$
give in fact
a set of generators in $H^{4-*}({\cal M}_g)$ that are usually taken as
follows \th\munozqu:
\eqn\gens{\eqalign{
\alpha &= 2 \mu (\Sigma_g) \in H^2 ({\cal M}_g), \cr
\psi_i&=\mu (\gamma_i) \in H^3 ({\cal M}_g), \cr
\beta&=-4 \mu (x) \in H^4 ({\cal M}_g), \cr}}
where $x$ is the class of the point in $H_0 (\Sigma_g)$.
We also define the ${\rm Sp}(2g, \IZ)$-invariant cohomology class in $H^6
({\cal M}_g)$,
\eqn\invt{
\gamma =-2 \sum_{i=1}^g \psi_i \psi_{i+g}.}
One can show that the moduli space of anti-self-dual connections on
$\Sigma_g \times {\bf S}^2$ with instanton number zero
is isomorphic to the moduli space of flat connections on $\Sigma_g$. In
particular, the
generators of the cohomology in \gens\ correspond precisely to the
Donaldson cohomology classes, and
we have that
\eqn\idents{
\alpha =2 I(\Sigma_g), \,\,\ \psi_i= I (\gamma_i),\,\,\, \beta=-4 {\cal O},}
while the invariant form $\gamma$ corresponds to \invform.

\subsec{The intersection pairings}
To determine the ring structure of the cohomology of ${\cal M}_g$, once a
set of
generators has been found, one only has to find a set of relations. Due to
Poincar\'e
duality, the intersection pairings of the generators in \gens\ give all the
information
needed to find the relations. In other words, to find the structure of the
cohomology ring
it is enough to evaluate the intersection pairings
\eqn\pairings{
\langle \alpha^m \beta^n \gamma^p \rangle_{{\cal M}_g}=
\int_{{\cal M}_g} \alpha^m \wedge \beta^n \wedge \gamma^p,}
as all the intersection pairings involving the $\psi_i$'s can be reduced
to \pairings\ by ${\rm Sp}(2r, \IZ)$ symmetry. These numbers can be
considered as the two-dimensional analogs of Donaldson invariants, which in
fact can
be formulated as correlation functions of a two-dimensional topological
gauge theory \revis\bt. In \th, Thaddeus computed \pairings\
in two steps: first, he obtained a recursive relation which allows
to eliminate the $\gamma$ classes. Second, he computed the pairings
$\langle \alpha^m \beta^n \rangle$
using Verlinde's formula \verlinde. The same steps are followed by
Witten in
\revis. We will also prove first the recursion relation,
and then compute the remaining pairings. To do this, we use the Donaldson
invariants
of product ruled surfaces. The relation between the pairings in \pairings\
and the Donaldson
invariants are as follows (see \munozqu, Remark 13):
\eqn\rel{
\langle \alpha^m \beta^n \gamma^p \rangle_{{\cal M}_g}=\epsilon([{\bf S}^2])
{\cal D}^{w_2=[{\bf S}^2]}_{\Sigma_g \times {\bf S}^2} \bigl(
(2\Sigma_g)^m(-4 x)^n \iota^p\bigr).}
In this equation, $\epsilon (w)=(-1)^{{K\cdot w + w^2 \over 2}}$, where $w$
is an integer
lift of the second Stiefel-Whitney class. This sign appears due to the
following reason.
The Donaldson invariants that appear in the right hand side are defined
using the natural
orientation of the moduli space of anti-self-dual connections. For
algebraic surfaces, this moduli
space can be realized as the Gieseker compactification of the moduli space
of rank two
stable bundles ${\cal M}(c_1,c_2)$,
where $c_1=w$. This complex space has a natural orientation induced by its
complex structure,
and the difference between these orientations in the computation of the
invariants
is given by $\epsilon (w)$. Now, when $c_2=0$
and $c_1=[{\bf S}^2]$ ({\it i.e.}, there is a non-zero flux through
the Riemann surface $\Sigma_g$), then ${\cal M}(0,c_1) = {\cal M}_g$,
and the intersection pairings in the left hand side are in fact computed
with the complex
orientation. In this case, $\epsilon([{\bf S}^2])=-1$. Notice that the
pairing above is only different from
zero when $2m+4n+6p=6g-6$. The Donaldson invariants in \rel\ can be computed
in any chamber, since for $c_2=0$ and $w_2(E)=[{\bf S}^2]$
one has $p_1=0$ and therefore there are no walls. The computation turns out
to be much simpler in the chamber of small volume for ${\bf S}^2$. One can
make in principle
the computation
in the other limiting chamber, using
$Z_{\Sigma_g}^{\epsilon=0}$. This turns out
to be very complicated
analytically, although we have explicitly checked
that the answers agree in many cases. Physically, the
computation of Thaddeus' intersection pairings
(as well as the computation of Donaldson invariants involving
low instanton numbers) is rather semiclassical
and is best performed in the electric frame,
while the ``magnetic" expression $Z_{\Sigma_g}^{\epsilon=0}$ gives
information about global aspects of the
generating function which are useful, for example, to compute
the eigenvalue spectrum of Fukaya-Floer cohomology.
It has also been noticed in \bjsv\ that in fact, to compute
the intersection pairings on ${\cal M}_g$,
the chamber of small volume for ${\bf S}^2$ is more natural, as the
topological reduction in this chamber gives
the twisted ${\cal N}=2$ Yang-Mills in two dimensions in
a direct way. We will then extract these pairings
from the Donaldson invariants given by \ruledexpf.

The first thing that we can prove is the
recursive relation of Thaddeus. Using
the explicit formula \ruledexpf, one easily sees that
\eqn\rec{
{\partial \over \partial r}Z^{w_2(E)=[{\bf S}^2]}_{g,{\bf S}^2}=2g
Z^{w_2(E)=[{\bf S}^2]}_{g-1,{\bf S}^2},}
and this implies, using \rel, that
\eqn\thred{
\langle \alpha^m \beta^n \gamma^p \rangle_{{\cal M}_g}=2g \langle \alpha^m
\beta^n
\gamma^{p-1} \rangle_{{\cal M}_{g-1}},}
which is precisely Thaddeus' recursive relation.

We now compute the intersection pairings $\langle \alpha^m \beta ^n \rangle$.
To do this, we use the expansion:
\eqn\expa{
\csc z = \sum_{k=0}^{\infty} (-1)^{k+1}(2^{2k}-2) B_{2k}{z^{2k-1} \over
(2k)!},}
where $B_{2k}$ are the Bernoulli numbers. We have to extract now the powers
$s^m p^n$ from the generating function \ruledexpf. Notice that a power $s^g$
comes already from the overall $g$-dependent factor in \ruledexpf. We then have
to extract the power $s^{m-g}$ from the series expansion in $s/2h$. taking
now into account
the comparison factors from \rel, and the dimensional constraint
$2m+4n=6g-6$, one finds
\eqn\computa{
\langle \alpha^m \beta ^n \rangle= {1 \over 4} 2^m (-4)^n i^{m-g+1}
2^{2g+n-m} m!
 { (2^{m-g+1}-2)\over
(m-g+1)!}  B_{m-g+1}[h_{\infty}^{3g-m-2}u_{\infty}^n f_{1\infty}^g
f^{-1}_{2 \infty}]_{q^0}.}
Fortunately, only the leading term contributes in the $q$-expansion
involved in \computa.
One finally obtains,
\eqn\thfor{
\langle \alpha^m \beta^n \rangle=(-1)^g { m! \over (m-g+1)!}2^{2g-2}
(2^{m-g+1}-2)  B_{m-g+1},}
which is exactly Thaddeus' formula for the intersection pairings.
This expression, as well as the recursive relation \thred, has been
obtained by Witten in \revis\ by exploiting the relation to physical
two-dimensional Yang-Mills theory. A derivation of \thfor\
based on gluing techniques has been worked out in \donglu.

One can formally consider the intersection pairings in the case of even
degree, and
extract them from the Donaldson invariants for vanishing Stiefel-Whitney
class. These invariants
are given by \ruledexp. Using now the expansion
\eqn\expcot{
\coth z = \sum_{k=0}^{\infty} 2^{2k} B_{2k}{z^{2k-1} \over (2k)!},}
and taking into account that $\epsilon(0)=1$, one finds
\eqn\thtfor{
\langle \alpha^m \beta^n \rangle=(-1)^g { m! \over (m-g+1)!} 2^{m+g-1}
B_{m-g+1}.}
These are also the pairings that one obtains using two-dimensional gauge theory
\revis\bt. However, one has to be extremely careful about the mathematical
interpretation
of \thtfor, since the space ${\cal M}_g^+$ is singular for $g \ge 3$. To
define the
intersection pairings one has then to use intersection cohomology \kiem\ or
consider a
partial desingularization of ${\cal M}_g$ that has only orbifold
singularities \hatter.
The computation of the intersection pairings for the partial
desingularization was
performed in \hatter\ using the strategy of \th. The result agrees with
\thtfor\ for
$m \ge g$, but for $m < g$ there are correction terms \foot{We are grateful
to Y.-H. Kiem
for explaining these issues to us.}. It would be interesting to see
if these corrections can be obtained using physical methods. In this sense,
the approach
in \revis\ seems more appropriate, as the singular character of the moduli
space
shows up as a non-regular term in the partition function.

\subsec{Relation to Verlinde's formula}
The derivation of the intersection pairings \thfor\ in \th\ was based in
the $SU(2)$
Verlinde's formula for the WZW model \verlinde. In fact, one can
reverse the logic in
\th\ and
give a derivation of Verlinde's formula from the intersection pairings. In
this section,
we will closely follow
the arguments given in \th\ for ${\cal M}_g$, and
we will also show that they can be formally extended
to ${\cal M}_g^+$.

Verlinde's formula gives an explicit expression for the number of
conformal blocks in CFT. In the case of the $SU(2)$ WZW
model, the space of conformal blocks at level $k$ (where $k$ is a
positive integer) can be identified
with the space of sections of the line bundle $L^{k/2}$, where $L$ is a
fixed line bundle
over ${\cal M}_g$ which generates ${\rm Pic}({\cal M}_g)\simeq \IZ$. The
canonical
bundle of ${\cal M}_g$ is given by $L^{-2}$. We are interested in
computing ${\rm dim} H^0 (L^{k/2},{\cal M}_g)$. As explained in \th, this
can be done
using Hirzebruch-Riemann-Roch. The canonical bundle of ${\cal M}_g$ is
negative, and by
the Kodaira vanishing theorem one has that $H^i (L^{k/2},{\cal M}_g)=0$ for
$i>0$. We then
have,
\eqn\hrr{
{\rm dim}\, H^0 (L^{k/2},{\cal M}_g)=\chi(L^{k/2}, {\cal M}_g)=
\int_{{\cal M}_g } {\rm ch}\, L^{k/2}\, {\rm td} \,{\cal M}_g.}
Notice that the cohomology classes involved in Hirzebruch-Riemann-Roch can be
expressed in terms of the generators of the cohomology ring \gens,
and therefore \hrr\ can be computed
in principle once the intersection pairings are known. Explicit expressions for
the characteristic classes of the tangent bundle to ${\cal M}_g$ have been
obtained by Newstead \news\ (see also \zag) and read
$c_1({\cal M}_g)=2\alpha$, $p({\cal M}_g)=((1+\beta)^{2g-2}$. One also has
$c_1(L)=\alpha$, and this gives:
\eqn\dim{
{\rm dim}\, H^0 (L^{k/2},{\cal M}_g)=
\int_{{\cal M}_g } \exp \Bigl( {k+2 \over 2} \alpha \Bigr) \biggl( {
{\sqrt \beta}/2 \over \sinh {\sqrt \beta} /2)} \biggr)^{2g-2}. }
If we expand the characteristic classes in the right hand side of \dim, we
get a polynomial in $k+2$
of the form
\eqn\dim{
\sum_{m=0}^{3g-3} {P_{3g-3-m}\over 2^{3g-3}} {(k+2)^m \over m!}
\int_{{\cal M}_g} \alpha^m \beta^{(3g-3-m)/2},}
where $P_n$ is the coefficient of $x^n$ in the series expansion of
$(x/\sinh x)^{2g-2}$. Using now
\thfor, it is easy to see that this is the coefficient of $x^{3g-3}$ in
\eqn\polyn{
\biggl( -{k+2 \over 2} x\biggr)^{g-1} \biggl({x \over \sinh
x}\biggr)^{2g-2} { (k+2)x \over \sinh (k+2)x}.}
An argument due to Zagier and explained in \th, Proposition (19), shows
that this coefficient is
given by
\eqn\verlodd{
{\rm dim}\, H^0 (L^{k/2},{\cal M}_g )=\biggl( {k+2 \over 2} \biggr)^{g-1}
 \sum_{n=1}^{k+1}
{ (-1)^{n+1} \over (\sin {n\pi \over k+2})^{2g-2}},}
which is precisely Verlinde's formula in the case of odd degree.

As we pointed out before, the moduli space of rank two stable bundles of
even degree is a singular space, and in principle the intersection numbers
are not
well-defined. The answer \thtfor\ should be considered as a regularization
of these
pairings in the context of the $u$-plane integral. If we assume that the
Riemann-Roch
formula is still valid, one obtains in fact the usual Verlinde's formula for
$SU(2)$ \foot{This has also been observed in \szenes\mohri.}. If
we consider the expression \dim\ with the intersection pairings
given in
\thtfor, one finds that
the dimension of $H^0 (L^{k/2}, {\cal M}_g^+)$ is now given by the
coefficient of $x^{3g-3}$
in
\eqn\polyndos{
-\biggl( -{k+2 \over 2}x\biggr)^{g-1} \biggl({x \over \sinh
x}\biggr)^{2g-2} { (k+2)x \coth (k+2)x}.}
Going through the argument in \th, Proposition (19), one easily obtains:
\eqn\verev{
{\rm dim}\,H^0 (L^{k/2},{\cal M}^+_g)= \biggl( {k+2 \over 2}\biggr)^{g-1}
\sum_{n=1}^{k+1} { 1 \over
(\sin {n\pi\over k+2})^{2g-2}},}
which gives the right formula for the number of conformal blocks for the
(untwisted)
$SU(2)$ case. This computation is, however, formal, as there is no suitable
Riemann-Roch
formula for a singular space like ${\cal M}^{+}_g$. Using the orbifold
desingularization of ${\cal M}^{+}_g$,
one can apply the Kawasaki-Riemann-Roch formula and relate the intersection
pairings
to Verlinde's formula \verev. In fact, this is how the corrections to the
pairings \thtfor\ are obtained
in \hatter.

\newsec{Application 2: Fukaya-Floer cohomology}

\subsec{Floer cohomology and gluing rules}

The Floer (co)homology groups of three-manifolds and their relations
to Donaldson invariants can be understood in a simple way using the
axiomatic approach
introduced by Atiyah \at, which in fact is a formalization of heuristic
considerations involving path integrals. According to the axiomatic approach,
a topological field theory in $3+1$-dimensions is essentially a functor
$\Phi$ from the
category of three-dimensional manifolds to the category of complex
vector spaces, $\Phi: {\rm Man}(3) \rightarrow {\rm Vect}$, and satisfying
certain
properties. In
the case of Donaldson-Witten theory, this functor associates to any
compact, oriented
three-manifold $Y$ the graded vector space given by the Floer homology groups
$HF_*(Y)$. These homology groups can be defined by using Morse theory with the
Chern-Simons functional on the moduli space of $SO(3)$ connections on $Y$
with second
Stiefel-Whitney class $w_2 \in H^2(Y, \IZ)$.

We will
be interested here in the gluing rules that relate the ring structure of
the Floer homology to the
Donaldson invariants of four-manifolds. Let us consider a four-manifold $X$
with boundary $\partial X =Y$, together with an element $z$ in ${\bf
A}(X)$. According to the
axiomatic approach, the functor $\Phi$ also assigns to the pair $(X,z)$
a ``relative invariant" of $X$, which is an element in the Floer
homology of $Y$ ({\it i.e.}, $\Phi(X,z) \in \Phi(\partial X)$).
This relative invariant can be understood in a simple way,
as explained in \tqft, in terms of path-integrals. Let $z=S_{i_1} \dots
S_{i_p}x^n \gamma_{j_1}
\dots \gamma_{i_q}$ be an element of ${\bf A}(X)$, where $S_{i_{\mu}}$,
$\gamma_{j_{\mu}}$ are two and
one-homology classes, respectively, and $x$ is the class of the point. This
determines a BRST-invariant
operator given by the $\mu$-map, namely
\eqn\opz{
{\cal A}_z = I(S_{i_1}) \dots I(S_{i_p}) {\cal O}^n I(\gamma_{j_1})
\dots I(\gamma_{i_q}).}
We then define the relative invariant through the usual correspondence
operators/states
in quantum field theory:
\eqn\rel{
(\Phi (X, z))^w_X(\phi_Y) = \int_{X} [D\phi]\bigl| _{\phi|_Y =\phi_Y}{\rm
e}^{-S_{\rm TYM}}{\cal A}_{z},}
which is a functional of the fields restricted to the boundary. In this
path integral, one integrates
over all the gauge fields with Stiefel-Whitney class
$w_X\in H^2(X, \IZ)$, where $w_X$ restricts to $w$ on $Y$.

There are some extra structures in $HF_*(Y)$ that will be important to our
analysis.
Our presentation will closely follow the excellent surveys in
\munozapp\munozfloer; for more details,
one can see \donaldsonii\floer. First of all, there is an
associative and graded commutative ring structure $HF_*(Y)\otimes HF_*(Y)
\rightarrow
HF_*(Y)$. Second, as in ordinary homology, one can define the dual of
$HF_*(Y)$ to
obtain the Floer cohomology of $Y$. Moreover, if $-Y$ denotes the manifold
with opposite
orientation, one has $HF^*(Y) \simeq HF_*(-Y)$. Finally, there is a
natural, non-degenerate
pairing ${\langle}\, ,\,{\rangle}:HF^*(Y) \otimes HF_{*}(Y) \rightarrow
\IC$. When the states
in the Floer cohomology are given by relative invariants, this pairing can be
understood heuristically from path integral arguments.
Consider two manifolds with boundary, $X_1$, $X_2$, such that
$\partial X_1=Y$ and $\partial X_2=-Y$ ({\it i.e.},
$Y$ with the opposite orientation). We can glue the manifolds together to
obtain a closed
four-manifold $X$. There is then a pairing between $HF_*(Y)$ and $HF_*(-Y)$
which is given
by
\eqn\pair{
\langle \Phi(X_1,z_1)^{w_{X_1}},  \Phi (X_2, z_2)^{w_{X_2}} \rangle =
\int_X [D\phi]
{\rm e}^{-S_{\rm TYM}}{\cal A}_{z_1}{\cal A}_{z_2}.}
In other words, the pairing is essentially given by Donaldson invariants of
the four-manifold $X$.

In order to give a precise gluing result, we have to be careful with two
things: first of all,
which is the Stiefel-Whitney class that one has to pick in order to define
the Donaldson
invariant on the right hand side of \pair? and, in case $b_2^+(X)=1$, in which
chamber should we compute the
Donaldson invariant? These issues are discussed in
\munozff\donaldsonii\ in detail. To answer
the first question, consider a $w \in H^2(X,\IZ)$ such that $w|_Y=w_2$.
Also consider a cohomology class
$[\Sigma] \in H^2(X, \IZ)$, given as the Poincar\'e dual of a
two-class
$\Sigma$ which
lies in the image of $i_*: H_2(Y, \IZ) \rightarrow H_2(X, \IZ)$, and
satisfying $w\cdot [\Sigma]=1$ (mod $2$).
The pair $(w, \Sigma)$ is called in \munozff\ an allowable pair.
One then defines
\eqn\definiti{
{\cal D}_X^{(w, \Sigma)}={\cal D}_X^w + {\cal D}_X^{w +
[\Sigma]}.}
The gluing theorem of \munozfloer\donaldsonii\ is then
\eqn\glupre{
\langle \Phi(X_1,z_1)^{w_{X_1}},  \Phi (X_2, z_2)^{w_{X_2}} \rangle
={\cal D}_X^{(w, \Sigma)}(z_1 z_2).}
If $b_2^+(X)=1$, then one considers the metric giving a long neck, {\it
i.e.}, one
takes $X=X_1\cup (Y \times [0,R])\cup X_2$, with $R$ very large.

We are interested in the Floer (co)homology of $Y=\Sigma_g \times {\bf
S}^1$, with
Stiefel-Whitney class $w_2 =[{\bf S}^1]$. This manifold
has an orientation-reversing diffeomorphism given by conjugation
on ${\bf S}^1$, and therefore there is a natural isomorphism $HF^*(Y) \simeq
HF_*(Y)$. We will
work with the ring cohomology from now on. The first thing to do is to
find the generators
of this ring. Consider the four-manifold with
boundary $X^0=\Sigma_g \times D^2$, where $D^2$ is the two-disk. Then,
$\partial X^0= \Sigma_g \times
{\bf S}_1$. We can define relative invariants of $X$ associated to
elements in ${\bf A}(X^0)={\bf A}(\Sigma_g)$. Clearly, one has to take $w
=[D^2] \in H^2(X^0, {\bf Z})$,
which restricts to $[{\bf S}^1]$ at the boundary. The generators of
$HF^*(Y)$ are
then given by \munozfloer:
\eqn\genfloe{\eqalign{
\alpha &= 2 \Phi^{w} (X^0, \Sigma_g) \in HF^2(Y), \cr
\psi_i&=\Phi^{w} (X^0,\gamma_i) \in HF^3(Y), \cr
\beta&=-4 \Phi^{w} (X^0,x) \in HF^4 (Y), \cr}}
where $\Sigma_g$, $\gamma_i$ and $x$ are the generators of $H_*(\Sigma_g)$.
Notice that this basis is very similar to the basis of ${\cal M}_g$
presented in \gens. In fact, $HF^*(Y)$ and $H^*({\cal M}_g)$ are isomorphic
as vector spaces \ds. The product structure in the Floer cohomology is given,
for these relative invariants, by $\Phi^w (X^0,z)\Phi^w (X^0,z')=\Phi^w
(X^0,zz')$.
We will restrict ourselves to the invariant part of $HF^*(Y)$ (as in the
analysis of
the cohomology of ${\cal M}_g$), which is generated by $\alpha$, $\beta$ and
\eqn\gam{
\gamma=-2 \sum_{i=1}^g \Phi^w (X^0, \gamma_i \gamma_{i+g}).}
The last ingredient we need is the gluing rule. If we consider the pairing
of two
relative invariants constructed from $X^0$, we will have to glue two copies of
$X^0$ along their boundaries. Clearly, this gives the closed four-manifold $X=
\Sigma_g \times {\bf S}^2$ (where ${\bf S}^2$ comes from gluing the two
disks along
their boundaries ${\bf S}^1$). The long neck metric is the one that makes
${\bf S}^2$
very big, and then corresponds to the chamber where $\Sigma_g$ is small.
Finally, we have
to specify the allowable pair. In $X$, $w=[{\bf S}^2]$ restricts to
$w_2=[{\bf S}^1]$ on
$Y$. On the other hand, the image of $H_2(Y,\IZ)$ in $H^2(X,\IZ)$ is
generated by $\Sigma_g$.
This means that the gluing rule for the relative invariants is
\eqn\glupart{
\langle \Phi^{w_0}(X^0,z_1),  \Phi ^{w_0}(X^0, z_2) \rangle
={\cal D}_X^{(w, \Sigma_g)}(z_1 z_2)={\cal D}_X^{w_2=[{\bf S}^2]}(z_1z_2) +
{\cal D}_X^{w_2=[{\bf S}^2]+[\Sigma_g]}(z_1z_2).}

The Fukaya-Floer cohomology $HFF^*(Y)$ of an oriented three-manifold $Y$
needs the extra input of a
loop $\delta \simeq {\bf S}^1$ in $Y$. A review of this construction can be
found in
\munozff. Here we will consider that $\delta$ is the
${\bf S}^1$ factor in $Y=\Sigma_g \times {\bf S}^1$.
In this case, one has that $HFF^*(Y)=HF^*(Y) \times \IC [[t]]$. A basis of
generators can be also
constructed using relative invariants of the manifold $X^0$, with the
insertion of the operator
$\exp t I(D^2)$ in the path integral. In this way, we obtain the generators
\eqn\genfloe{\eqalign{
\widehat\alpha &= 2 \Phi^{w_0} (X^0, \Sigma_g {\rm e}^{t D^2}) \in
HFF^2(Y), \cr
\widehat\psi_i&=\Phi^{w_0} (X^0,\gamma_i{\rm e}^{t D^2}) \in HFF^3(Y), \cr
\widehat\beta&=-4 \Phi^{w_0} (X^0,x {\rm e}^{t D^2}) \in HFF^4 (Y). \cr}}
The gluing rule is now
\eqn\glufu{
\langle \Phi^{w_0}(X^0,z_1 {\rm e}^{t D^2}),  \Phi ^{w_0}(X^0, z_2{\rm
e}^{t D^2}) \rangle
={\cal D}_X^{(w, \Sigma_g)}(z_1 z_2 {\rm e}^{t {\bf S}^2}),}
and therefore the Donaldson invariants involved in the Fukaya-Floer cohomology
include
the cohomology class associated to ${\bf S}^2$. This makes the determination of
this cohomology more difficult.

\subsec{Eigenvalue spectrum of the Fukaya-Floer cohomology}
As we have seen, the intersection pairings in Floer and Fukaya-Floer
cohomology are given by Donaldson invariants of $\Sigma_g \times {\bf S}^2$
in the chamber where $\Sigma_g$ is small. These invariants
completely determine, in principle, the ring structure of the
(Fukaya)-Floer cohomology,
but this does not mean that we are able to give an explicit presentation of the
relations of the ring. Already in the comparatively simpler case of
the classical cohomology of ${\cal M}_g$, to obtain the explicit relations
at genus $g$
starting from the intersection pairings \thfor\ turns out to be a very
complicated combinatorial problem (solved in \zag). In this section, we
want to show that
an important aspect of the ring structure,
namely the eigenvalue spectrum, can be deduced in a simple way from the
generating
function that we found in section 2. In the case of Floer cohomology, the
spectrum was
obtained in \bjsv\ under some extra assumptions, and finally derived in
\munozfloer\
from an explicit presentation of the relations. The spectrum of the
Fukaya-Floer cohomology
was conjectured in \munozff, based on the computation of the spectrum for a
submodule. Our
calculation confirms this conjecture. Our strategy will be in a way the
reverse to that in
\munozapp. In these papers, the information about Fukaya-Floer cohomology
obtained in \munozff\ is used to
understand the structure of Donaldson invariants. Here, we will use the
Donaldson
invariants of $\Sigma_g \times {\bf S}^2$ to deduce results about the
Fukaya-Floer cohomology of $\Sigma_g \times {\bf S}^1$.

The basic procedure to obtain the eigenvalue spectrum is to find
elements in the ideal of relations of the Fukaya-Floer cohomology, {\it i.e.},
to find vanishing polynomials in the generators $\widehat \alpha$,
$\widehat\beta$ and
$\widehat\gamma$:
\eqn\poly{
{\cal P}(\widehat \alpha , \widehat \beta, \widehat \gamma)=0.}
We can easily translate this identity in terms of the generating function
for the Donaldson invariants of $\Sigma_g \times {\bf S}^2$: as the pairing
\pair\ is non-degenerate, to prove the identity \poly\ it is enough to prove
that
\eqn\pairpol{
\langle {\cal P}(\widehat\alpha,\widehat\beta, \widehat\gamma),
\Phi (X^0, z{\rm e}^{t D^2}) \rangle =0,}
for any $z \in {\bf A}(\Sigma_g)$. Due to the gluing rule \glufu, the above
pairing is nothing
but ${\cal D}_X^{(w, \Sigma_g)}({\cal P}(2\Sigma_g, -4x, \iota)z {\rm
e}^{t {\bf S}^2})$.
The vanishing of \pairpol\ for any $z$ is then equivalent to the following
differential
equation,
\eqn\diffeq{
{\cal P} \bigl( 2 {\partial \over \partial s}, -4 {\partial \over \partial
p}, {\partial
\over \partial r} \bigr) Z_{g}^{(w, \Sigma_g)}(p,r,s,t)=0,}
where we have defined the generating functional corresponding to the invariants
\definiti:
\eqn\genpair{
Z_g^{(w, \Sigma_g)}(p,r,s,t)=Z_g^{\epsilon=0}(p,r,s,t) +
Z_g^{\epsilon=1}(p,r,s,t).}
What we have computed in section 3 are precisely the generating
functions involved in \genpair. We then have to study the differential
equations satisfied by our function. First of all, using \uplanetwo\ and
\switten, one immediately finds:
\eqn\genpairex{
\eqalign{
 Z_g^{(w, \Sigma_g)}=
&-2^9 i^{1-g}
\Bigg[ {\rm e}^{-2pu_M-2st T_M}
\sum_{m=1}^g {g \choose m}(-1)^m 2^{-6m}(\tilde h_M^2\tilde f_{2M})^{m-1}\cr
&\,\,\,\,\,\,\  \quad \quad \quad \cdot
\biggl( 2ir - {is \over 16} \tilde h_M^2 \tilde f_{1M} \biggr)^{g-m} {\rm
Li}_{-m}
(-{\rm e}^{-2it\tilde h_M^{-1}})\Bigg]_{q_D^0} \cr
& +2^{9}
\sum_{\scriptstyle a \,{\rm odd}\atop \scriptstyle 0<ab\leq g-1 } \Bigg[
{\rm sgn (a)} q_D^{ab}(-1)^{b}\ex^{2p u_M+ 2 s t T_M} (\tilde h_M^2\tilde
f_2)^{-1}\cr
&\cdot \Biggl(
\biggl( {s\over 16 } \tilde h_M^2 \tilde f_{1M}
+2^{-7} b \tilde h_M^2\tilde f_{2M}+ 2r \biggr)^g \ex^{{2as-2bt\over \tilde
h_M}} \Biggr)
\Bigg]_{q_D^0}\cr
& +2^{9}i^{1-g}
\sum_{\scriptstyle a \,{\rm even}\atop \scriptstyle 0<ab\leq g-1
}\Bigg[{\rm sgn (a)}
q_D^{ab}(-1)^{b}\ex^{-2pu_M-2 s t T_M} (\tilde h_M^2\tilde f_2)^{-1}\cr
&\cdot \Biggl(
\biggl( -{i s\over 16 }\tilde h_M^2 \tilde f_{1M}
+ 2^{-7} b \tilde h_M^2 \tilde f_{2M}+ 2i r \biggr)^g
\ex^{-{2ias-2i b t\over \tilde h_M}} \Biggl)
\Bigg]_{q_D^0}.\cr}}
This generating function can be explicitly evaluated for low genus. The results
are specially simple if we put $t=0$. \foot{For $t=0$, the generating
function \genpairex\ can be computed in principle using the Artinian
decomposition of the
Floer cohomology \munozfloer. This procedure, however, does not give a
general result for any genus and has to be worked out case by case.
V. Mu\~noz has informed us that the above
expressions
for $g=2,3$ coincide with the results that can be obtained from this
decomposition.}
We obtain, for example:
\eqn\examples{
\eqalign{
Z_1^{(w,\Sigma_1)}=&-{\rm e}^{-2p},\cr
Z_2^{(w, \Sigma_2)}(p,r,s)=& -{ 1 \over 8}{\rm e}^{-2\,p}\left( 32\,r - s
\right)
        -{1\over 32} {\rm e}^{2p}({\rm e}^{2s}-{\rm e}^{-2s}),\cr
Z_3^{(w, \Sigma_3)}(p,r,s)= &-{1 \over 4096}{\rm e}^{-2p}
\left( 98 + 256\,p + 256\,{p^2} + 49152\,{r^2} - 3072\,r\,s +
       48\,{s^2} \right)  \cr
&+ {1\over 256}{\rm e}^{2p +2s} (3 - 4\,p - 48\,r - 2\,s) +
{1\over 256}{\rm e}^{2p -2s} (3 - 4\,p + 48\,r + 2\,s) \cr&+{1 \over 4096}
{\rm e}^{-2p} ({\rm e}^{4is} +{\rm e}^{-4is}).
\cr     }       }

Let us now concentrate on the eigenvalue spectrum of the Fukaya-Floer
cohomology. The first eigenvalue equation we can write is the one that
corresponds
to the finite-type condition that we obtained in section 3. It reads now,
\eqn\finibet{
(\widehat \beta^2 - 64)^g=0,}
therefore the eigenvalues of $\widehat\beta$ must be $\pm 8$. To
understand the eigenvalue
spectrum of the remaining operators, it is useful first to be more precise
about
the structure of $Z_g^{(w, \Sigma_g)}$. If we look at \genpairex, it is easy to
see that it can be written as,
\eqn\struct{
Z_g^{(w, \Sigma_g)}=\sum_{|a|\le g-1} Z_a (p,r,s,t).}
Notice that the $u$-plane integral contribution corresponds to $a=0$. The
structure of
$Z_a (p,r,s,t)$ is immediate from \genpairex:
\eqn\za{
Z_a (p,r,s,t)=\cases{f_a(p,r,s,t) {\rm e}^{2p +st + 2as}, & for $a$ odd,\cr
              f_a(p,r,s,t) {\rm e}^{-2p -st - 2i as},& for $a$ even,} }
where $f_a(p,r,s,t)$ is a polynomial in $p$, $r$ and $s$ and a power series
in $t$ (similar
remarks about the structure of the Donaldson invariants have been made in
\munozapp).
We are interested in the degree of the polynomial in $p$, $r$, $s$. This is
again easy to
see if we look at the modular forms involved in \genpairex. Assume $g\ge 2$
(for $g=1$,
$f_0$ only depends on $t$). We know that the maximum power we can find in
$p$ is precisely
$g-1$ (a simple consequence of the finite type condition). As one can see
in \menosun,
this power appears in the $u$-plane integral contribution, and for
$a\not=0$, the maximum power of
$p$ is in fact $g-2$.

Let us now find which is the maximum power
of $s$ in $f_a$. If we group the powers of $s$ in the modular form that
gives $f_a$, we easily see that
the leading term in $q_D$ has the form
\eqn\powers{
q_D^{ab+1+n-m} s^{g+n-m} t^n,}
up to numerical constants. It is clear that the maximum possible power of $s$
which can appear in $f_a$ is $g-|a|-1$, and occurs  for $|b|=1$. This power
actually appears in $f_a$: using the above expression, it is easy to see that
\eqn\mapodd{
f_a(p,r,s,t)=- 2^{6-4g-3|a|}\sum_{n=|a|+1}^g {g \choose n}({\rm
sgn}(a))^{n+1}{(a+2t)^{n-|a|-1} \over (n-|a|-1)!}
{\rm e}^{-2({\rm sgn}(a)) t} s^{g-|a|-1} + \dots,}
for $a$ odd, while for $a$ even one obtains:
\eqn\mapev{
\eqalign{
&f_a(p,r,s,t)=\cr&(-1)^{g-|a|} i^{|a|}2^{6-4g-3|a|}\sum_{n=|a|+1}^g {g
\choose n}({\rm sgn}(a))^{n+1}
{ (a-2it)^{n-|a|-1} \over (n-|a|-1)!}
{\rm e}^{2i({\rm sgn}(a)) t} s^{g-|a|-1} + \dots.\cr}}
Finally, for $a=0$ (the $u$-plane contribution), one has:
\eqn\mapze{
f_a(p,r,s,t)=(-1)^g 2^{6-4g}\sum_{n=1}^g { (-1)^n (-2it)^{n-1} \over (n-1)!}
 {g \choose n}{\rm Li}_{-n}(-{\rm e}^{-2it})
 s^{g-1} +
\dots.}
Notice that, for $a\not=0$, $f_a(p,r,s,t)$ is a polynomial in ${\rm e}^{\pm
t}$, while for $a=0$
it is a rational function in these variables.

By similar arguments, one finds that the maximum power of $r$ appears for $a=0$
and is $g-1$. We then have the following differential equations for the
$Z_a$:
\eqn\morediff{
\eqalign{
\bigl( -4 {\partial \over \partial p} + 8\bigr)^{g-1} Z_a=
\bigl(2 {\partial \over \partial s} -4a-2t\bigr)^{g-|a|}Z_a&=0, \,\,\,\,\,\
a\,\,\,\, {\rm odd} \cr
\bigl( -4 {\partial \over \partial p} - 8\bigr)^{g-1} Z_a=
\bigl(2 {\partial \over \partial s} +4ia +2t \bigr)^{g-|a|}Z_a&=0,
\,\,\,\,\,\ a\,\,\,\, {\rm even}, \,\,\,\ a\not=0, \cr
\bigl( -4 {\partial \over \partial p} - 8\bigr)^{g} Z_0=
\bigl(2 {\partial \over \partial s} +2t \bigr)^{g}Z_0&=0,\,\,\,\,\,\ a=0 }.}
Notice that the powers that appear in these equations are in fact the
minimum powers that
are needed to kill the
$Z_a$, as it can be easily seen from \mapodd\mapev\ and \mapze. We also have
\eqn\moreg{
{\partial^g \over \partial r^g}Z_g^{(w, \Sigma_g)}=0,}
which is also the minimum power we need ($r^{g-1}$ appears for $a=0$, while for
$a \not=0$ the maximum power is $r^{g-2}$).
Notice, in particular, that $(\widehat \beta \pm 8)^g$ kills all the $Z_a$
with $a$
odd (even, respectively).
We can now deduce the eigenvalue spectrum of the Fukaya-Floer cohomology. From
\morediff\ and \moreg\ we find the
following operator equations:
\eqn\nilp{
\eqalign{
\widehat \gamma^g=&0, \cr
\prod_{a \, {\rm odd}} (\widehat\alpha -4a-2t)^{g-|a|} \prod_{a \, {\rm even}}
(\widehat\alpha +4ia+2t)^{g-|a|}=&0.\cr}}
Therefore, the only eigenvalue of $\widehat \gamma$ is $0$, and for
$\widehat \alpha$ we find
the eigenvalue spectrum $(0,\pm 4 +2t, \pm 8i -2t, \cdots)$. Notice that
the eigenvalue $8$
of $\widehat \beta$ only occurs for $\widehat \alpha =-4ia-2t$, with $a$ even.
In the same way, we find that $-8$ only occurs for $\widehat \alpha =4a+2t$,
for $a$ odd. This is due to the
equation
\eqn\apar{
(\widehat \beta+8)^g \prod_{a \, {\rm even}} (\widehat\alpha
+4ia+2t)^{g-|a|}=0,}
and a similar equation involving $(\widehat \beta-8)^g$. Recalling now that
$-(g-1) \le a \le g-1$, our main conclusion is that the
eigenvalue spectrum of $(\widehat \alpha,\widehat \beta, \widehat \gamma)$
is given by
\eqn\spec{
(0,8,0), \,\,\,\ (\pm 4 +2t, -8,0), \,\,\,\ \dots \,\,\,\
(\pm 4(g-1)i^g + (-1)^g 2t, (-1)^{g-1} 8,0).}
This generalizes Proposition 20 in \munozfloer, and
confirms the conjecture in \munozff\ (see Theorem 5.13 and Remark 5.14 in
that paper).
It is easy to give an explicit construction of the eigenvectors
corresponding to these eigenvalues.
They are given by:
\eqn\eigenv{
v_a=\cases{\bigl(\widehat \beta + 8\bigr)^g
\bigl(\widehat \alpha +4ia +2t \bigr)^{g-|a|-1}
\prod_{\scriptstyle a'\,{\rm even},\, a'\not=a \atop \scriptstyle |a'|\leq g-1}
\bigl(\widehat \alpha +4ia' +2t \bigr)^{g-|a'|}, &$a$ even,\cr
\bigl( \widehat \beta - 8\bigr)^g
\bigl(\widehat \alpha -4a -2t \bigr)^{g-|a|-1}
\prod_{\scriptstyle a'\,{\rm odd},\, a'\not=a \atop \scriptstyle |a'|\leq g-1}
\bigl(\widehat \alpha -4a' -2t \bigr)^{g-|a'|},& $a$ odd.}}
To see that these vectors are in fact not zero, one can easily prove that
$\langle v_a,
\Phi (X^0, z{\rm e}^{t {\bf S}^2}\rangle
\not= 0$ for any $z \in {\bf A}(X)$, using the explicit results for
the generating function \genpairex\ in \mapodd, \mapev\ and
\mapze.

Using now arguments from \bjsv\munozfloer\sieb, together with our
results, it is
easy to rederive the presentation of the Floer cohomology of $\Sigma_g
\times {\bf S}^1$
given in \bjsv\munozfloer. We will give some brief indications on this respect.
Let $J_g$ be the ideal
of relations at genus $g$ ($J_g$ is then generated by all
the polynomials in $\alpha, \beta, \gamma$ that
vanish as elements of the invariant part of $HF^*$). First of all, notice that
$Z_g^{(w, \Sigma_g)}$ satisfies
\eqn\recurs{
{\partial \over \partial r} Z_g^{(w, \Sigma_g)}=2g Z_{g-1}^{(w, \Sigma_g)},}
This implies immediately the following inclusion relation:
\eqn\rec{
\gamma J_g \subset J_{g+1} \subset J_g.}
Now one can use the fact that the Floer cohomology of $Y=\Sigma_g \times
{\bf S}^1$ is a deformation of the
cohomology of ${\cal M}_g$ (this is rather elementary and does not assume
the existence of a ring isomorphism between $HF^{*}(Y)$ and the quantum
cohomology of ${\cal M}_g$). Using the explicit recursive presentation of
the ring cohomology of ${\cal M}_g$
given in \zag\st, and adapting the
arguments of Proposition 3.2 in \sieb, one obtains the following result:
the ideal of relations is given by
$J_g=(q^1_g,q^2_g, q^3_g)$, where the $q^i_g$, $i=1,2,3$, are given
by the following recursive relations:
\eqn\recur{
\eqalign{
q^1_{g+1} &=\alpha q^1_g + g^2 q_g^2,\cr
q^2_{g+1} &=(\beta + c_{g+1})q^1_g + {2g \over g+1} q_g^3,\cr
q^3_{g+1} &=\gamma q^1_g.\cr}}
When $c_{g+1}=0$, one recovers the classical cohomology
of ${\cal M}_g$. The deformation is then encoded in the coefficient $c_{g+1}$.
Notice that the key fact which is used to derive \recur\ is the inclusion
of ideals \rec, which is
in turn a consequence of \recurs.
This recursion relation was conjectured in \sieb\ for the generating
function of the Gromov-Witten
invariants of ${\cal M}_g$, and provides in that context a generalization of
Thaddeus' recursion relation \thred\
from the classical to the quantum pairings.

The last ingredient is then to compute the value of the coefficient $c_{g+1}$.
It was shown in \munozfloer\ that this value can be easily deduced by induction
using the eigenvector that corresponds to the maximum $\alpha$-eigenvalue.
For $g=1$,
one immediately finds from \genpairex\ that
$Z_1^{(w, \Sigma_1)}=-{\rm e}^{-2p}$ (we are putting $t=0$, as we are
considering
the Floer rather than the Fukaya-Floer cohomology). It then follows
that $\beta =8$, so $c_1=-8$. The argument in \munozfloer, Theorem 14,
gives then $c_{g}=(-1)^g 8$.

It would be interesting to use the information contained in \genpairex\ to
give a recursive presentation like \recur\ but for the Fukaya extension. Some
steps in this direction have been taken in \munozff. In principle, all the
information that one needs is contained in the generating function
\genpairex, but it is
still a non-trivial problem to extract it in the particular form of an explicit
presentation of the relations.

\bigskip
\centerline{\bf Acknowledgements}
\bigskip

We would like to thank G. Moore for explanations of \mw, for many
useful discussions on the topics treated in this paper, and for a critical
reading of the manuscript. We are
grateful to V. Mu\~noz for discussions and correspondence on Floer
cohomology and for communicating us his computations of generating
functions. We would also like to thank Y.-H. Kiem and B. Siebert
for useful discussions and correspondence. Finally, we want to thank
J.M.F. Labastida for a critical reading of the manuscript and for
his encouragement. C.L. would like to thank
the Physics Departments of Yale University and Brandeis University
for their hospitality. M.M. would like to thank the Departamento de
F\'\i sica de Part\'\i culas at the Universidade de Santiago de
Compostela, as well as the organizers of VBAC99, for their
hospitality during the completion of this paper. The work of C.L.
is supported by DGICYT under grant PB96-0960. The work of M.M. is
supported by DOE grant DE-FG02-92ER40704.

\listrefs
\end